\documentclass[twocolumns]{article}
\usepackage[utf8]{inputenc}
\usepackage[left=0.5in, right=0.5in, top=0.75in, bottom=0.75in]{geometry}
\usepackage{dblfloatfix}
\usepackage{graphicx}
\usepackage{caption}
\usepackage{xcolor}
\usepackage{microtype}
\usepackage{amsmath,amsfonts,amssymb}
\usepackage[symbol]{footmisc}
\usepackage{chemist}
\usepackage[hidelinks]{hyperref}
\usepackage{titlesec}
\usepackage[sort&compress,numbers]{natbib}
\usepackage[british]{babel}
\titleformat*{\section}{\large\bfseries}
\titleformat*{\subsection}{\normalsize\bfseries}

\begin{document}

\twocolumn[{
    \noindent\makebox[\linewidth]{\rule{\linewidth}{1.5pt}}
    
    \vspace{0.0in}
    \begin{center}
        {\Large \textbf{Emergent microrobotic oscillators via asymmetry-induced order}}
    \end{center}
	
	\vspace{-0.1in}
	
	\noindent\makebox[\linewidth]{\rule{\linewidth}{1.5pt}}
	
	\vspace{0.15in}

    \textbf{Jing~Fan~Yang}$^{*1}$ \ \ \ 
    \textbf{Thomas~A.~Berrueta}$^{*2}$ \ \ \ 
    \textbf{Allan~M.~Brooks}$^1$ \ \ \
    \textbf{Albert~Tianxiang~Liu}$^{1,3}$ \ \ \ 
    \textbf{Ge~Zhang}$^1$ \ \ \
    \textbf{David~Gonzalez-Medrano}$^4$ \ \ \
    \textbf{Sungyun~Yang}$^1$ \ \ \
    \textbf{Volodymyr~B.~Koman}$^1$ \ \ \
    \textbf{Pavel~Chvykov}$^5$ \ \ \
    \textbf{Lexy~N.~LeMar}$^1$ \ \ \
    \textbf{Marc~Z.~Miskin}$^4$ \ \ \
    \textbf{Todd~D.~Murphey}$^2$ \ \ \
    \textbf{Michael~S.~Strano}$^{\dagger1}$
    
    \vspace{0.05in}
    
    {\small $^1$Department of Chemical Engineering, Massachusetts Institute of Technology, Cambridge, MA, USA.\\
    $^2$Center for Robotics and Biosystems, Northwestern University, Evanston, IL, USA.\\
    $^3$Department of Chemical Engineering, University of Michigan, Ann Arbor, MI, USA.\\
    $^4$Department of Electrical and Systems Engineering, University of Pennsylvania, Philadelphia, PA, USA.\\
    $^5$Physics of Living Systems, Massachusetts Institute of Technology, Cambridge, MA, USA.}

    \vspace{0.05in}
}]

\footnotetext{$^*$These authors contributed equally to the work.}
\footnotetext{$^{\dagger}$Corresponding author: strano@mit.edu}

%%%%%%%%%%
% Abstract
\textbf{\small Spontaneous oscillations on the order of several hertz are the drivers of many crucial processes in nature. From bacterial swimming to mammal gaits, converting static energy inputs into slowly oscillating power is key to the autonomy of organisms across scales. However, the fabrication of slow micrometre-scale oscillators remains a major roadblock towards fully-autonomous microrobots. Here, we study a low-frequency oscillator that emerges from a collective of active microparticles at the air-liquid interface of a hydrogen peroxide drop. Their interactions transduce ambient chemical energy into periodic mechanical motion and on-board electrical currents. Surprisingly, these oscillations persist at larger ensemble sizes only when a particle with modified reactivity is added to intentionally break permutation symmetry. We explain such emergent order through the discovery of a thermodynamic mechanism for asymmetry-induced order. The on-board power harvested from the stabilized oscillations enables the use of electronic components, which we demonstrate by cyclically and synchronously driving a microrobotic arm. This work highlights a new strategy for achieving low-frequency oscillations at the microscale, paving the way for future microrobotic autonomy.}

%%%%%%%%%%%%%%
% Introduction
\section*{Introduction}
\label{sec:intro}
The ability to produce low-frequency oscillations is central to the autonomy of living beings, and is essential to key biological processes such as heartbeats, neuron firings, breathing, and locomotion~\cite{Buzsaki2004,Kruse2005,Katz2016}. While complex electronics operate at ever-increasing clock rates of many gigahertz, the frequency of many important biological oscillations seldom exceeds 100Hz. The slow rate of these oscillations stems from a need to be commensurate with both the energy budget and the natural timescales of underlying biological processes, as in the transport of CO$_2$ in plants~\cite{Minguet-Parramona2016} and in the galloping of horses~\cite{Hoyt1981}. Unlike oscillations arising from external periodic forcing~\cite{Yao1996optoelectronic,White2008,Lagzi2010,Gardi2022}, these self-oscillations emerge spontaneously from the balancing of competing dynamical processes driving systems away from equilibrium~\cite{Jenkins2013,Hua2021,He2012synthetic}---a signature of living systems~\cite{Grzybowski2016}. 

In artificial microsystems, however, the production of slow self-sufficient self-oscillations is counterintuitively difficult~\cite{Akbar2021,Shen2021}. Generating self-sustaining mechanical oscillations at the microscale typically requires the transduction of complex chemical oscillators (e.g., Belousov-Zhabotinsky reaction~\cite{Hudson1981}) into periodic changes to a system's physical configuration~\cite{Lagzi2010,Maeda2007,Altemose2017,Zhou2020,Yoshida2010,Onoda2017}. Alternative mechanisms for producing self-sufficient mechanical oscillations based on carefully designed dynamic coupling between responsive elastic materials and thermal~\cite{Zhao2019_hydrogel,He2012synthetic}, chemical~\cite{Hua2021,He2012synthetic,Horvath2011}, or moisture stimuli~\cite{Shin2018_SR} have typically been demonstrated in millimetre-scale (and larger) devices. In contrast, generating slow periodic electrical signals remains prohibitively challenging aboard untethered microscale devices (Supplementary Note 3), given the limited downward scalability of capacitors and inductors~\cite{Funaki2021,Molnar2021}, as well as the power and footprint demands of CMOS oscillators, frequency dividers, and energy modules~\cite{Funke2016,Hwang1995,Galea2018}. Despite these challenges, recent progress has shown that self-sustaining electrical oscillations can be produced by modulating electrical resistance with mechanical feedback loops in carefully designed devices, presenting a promising mechanism for sub-500$\mu$m electrical self-oscillators~\cite{Akbar2021}.

In this work, instead of relying on complex chemistries, integrated electronics, or elaborate mechanical microstructures, we produce robust electromechanical oscillations aboard a collective of deceptively simple microparticles by exploiting the self-organized properties of their far-from-equilibrium dynamics. By breaking the permutation symmetry of a homogeneous particle collective situated at an air-liquid interface, we reliably control their dynamics to realize simultaneous chemomechanical and electrochemical periodic energy transduction. We achieve this by introducing a particle with an enhanced reaction rate, whose stabilizing effect on the system behaviour we analyze through the lens of asymmetry-induced order. In turn, through a simple bimetallic on-board fuel cell design, we transduce the system's self-oscillations into periodic electrical work to power state-of-the-art microrobotic components, without the need for batteries or external sources of energy.
 
%%%%%%%%%%
% Figure 1
\begin{figure*}[pt]
    \centering
    \includegraphics[width=0.78\linewidth]{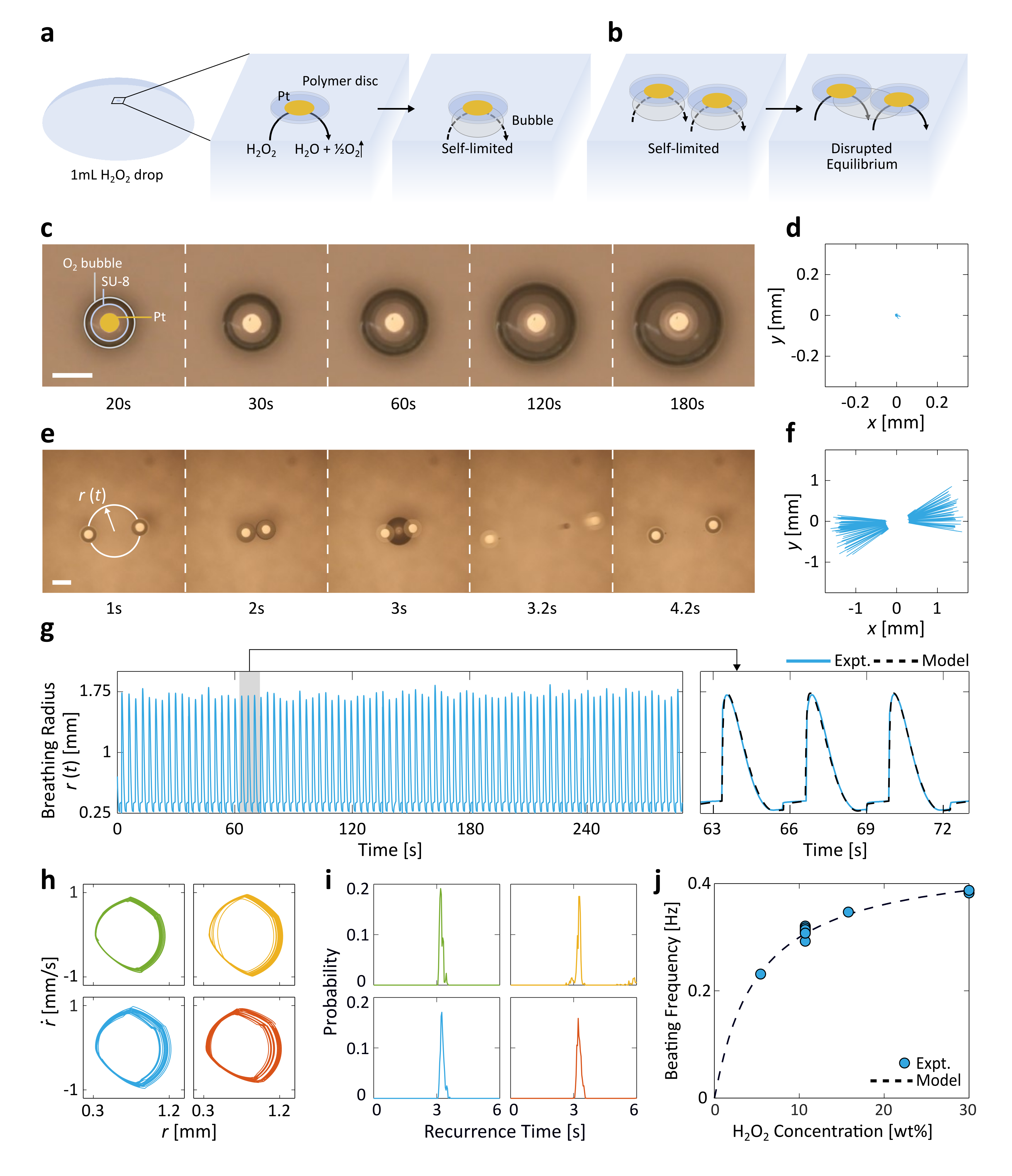}
    \caption{\small \textbf{Emergence of chemomechanical microparticle self-oscillation.}
    \textbf{a}, Schematic of a self-limited system of a single particle resting still at the air-liquid interface of a H$_2$O$_2$ drop. The particle is composed of a catalytic patch of Pt (yellow) underneath a polymeric disc (blue). The O$_2$ formation slows down asymptotically over time as the gas bubble restricts the available catalytic surface area. 
    \textbf{b}, A 2-particle system, in contrast, exhibits an emergent and self-sustained beating behaviour as the bubble merger restores the previously hindered reactivity, thus disrupting the equilibrium state.
    \textbf{c},\textbf{d}, Micrograph sequence (\textbf{c}) and tracked particle coordinates (\textbf{d}) of a 1-particle system that remains still for an extended period of time.
    \textbf{e},\textbf{f}, Micrograph sequence (\textbf{e}) and tracked coordinates (\textbf{f}) of a 2-particle system with emergent beating. The breathing radius, $r(t)$, is the distance from the collective's centroid to each particle, averaged over all particles.
    \textbf{g}, The long-term breathing radius trajectory of the same system as in (\textbf{e}) and (\textbf{f}) demonstrates the robustness of the beating behaviour. The shaded portion is magnified in the right panel, where the mechanistic model simulations (black, Supplementary Note 1) are shown to match the experimental curve (blue).
    \textbf{h}, The phase portraits of 4 independent 2-particle experiments demonstrate reproducible limit cycles with closed-loop orbits, confirming the periodicity of collective beating. Note that to calculate the phase portraits the system's bubble-driven discontinuities were processed through a standard finite-impulse response filter (see Methods). All phase portraits share the same axes.
    \textbf{i}, The recurrence histograms of the same 4 experiments all display a narrow peak centred at a period of 3.2s, consistent with visual evidence in (\textbf{e}). All histograms share the same axes.
    \textbf{j}, The beating frequency can be tuned with the concentration of H$_2$O$_2$. The dependence predicted by the mechanistic simulations on the basis of a Langmuir-Hinshelwood kinetics (black curve) matches the experimental measurements (blue markers).
    Scale bars, $500\mu$m.}
    \label{fig:fig1}
\end{figure*}

%%%%%%%%%
% Results
\section*{Results}
\label{sef:results}

%%%%%%%%%%%
% Section 1
\subsection*{Emergent low-frequency oscillation}
\label{sec:sec1}
Figure~\ref{fig:fig1} presents a system of simple microparticles where low-frequency chemomechanical self-oscillations emerge from the coupling of otherwise self-limiting catalytic reactions easily trapped at equilibrium. Figure \ref{fig:fig1}a shows that each of these microparticles, composed of nothing more than a nanometre-thick Pt patch of radius $125\mu$m microfabricated beneath a polymeric microdisc, generates a gas bubble when placed at the curved air-liquid interface of a H$_2$O$_2$ drop via 
\begin{chemmath}
  H_2O_2 \reactrarrow{0pt}{1cm}{\small Pt}{} H_2O+\frac{1}{2}O_2.
  \tag{1}\label{eq:peroxide_decom}
\end{chemmath}

\noindent This well-studied catalytic reaction has been a long-time favourite in both micro-~\cite{Wang2006_nanomotor,Paxton2006,Brooks2019,Bandari2020} and macroscopic robotics~\cite{He2012synthetic,Wehner2016}, noted for the fuel’s high energy density and simple chemistry~\cite{Wehner2016}.

For a single microparticle situated at the interface, the chemical reaction in Fig.~\ref{fig:fig1}a is self-limiting as the bubble grows and gradually blocks off the fuel’s access to the catalyst. Consequently, the single-particle system reaches its equilibrium state promptly: The microparticle remains motionless for a prolonged time (Fig.~\ref{fig:fig1}d, Supplementary Movie 1) and the bubble asymptotically reaches a terminal radius without rupture (Fig.~\ref{fig:fig1}c). However, a drastic change occurs when a second identical particle is introduced to the system. Figure~\ref{fig:fig1}b shows that as the microparticles enter each other's proximity, the separately-formed gas bubbles merge. The freed-up catalytic surface area then disrupts the self-limiting chemistry, destabilizing the original single-particle steady-state. This allows the merged bubble to grow beyond its threshold, leading to its rupture (Fig.~\ref{fig:fig1}e, $t=3.2$s). The collapse imparts an impulse onto the microparticles and propels them in opposite directions, at which point the particles are drawn back towards one another by the restorational forces: First, the radial component of buoyancy, $\mathbf{F}_\text{g}$, globally directs the particles towards the apex of the concave air-liquid interface~\cite{Gardi2022}. Second, the local interfacial deformations result in a mutual attractive capillary force $\mathbf{F}_{\text{c}}$, affectionately known as the ``Cheerios effect''~\cite{Vella2005,Xie2022_cargo}. The combination of this Cheerios effect and catalytic bubble generation has been observed to produce repetitive back-and-forth motion~\cite{Mei2008_functionalizedtubes,Solovev2009_jetengine} in swarms of tubular swimmers~\cite{Solovev2010_microstrider,Solovev2013_collective}.
All of these factors sum up to a repeatable cycle of mutual approach, contact, bubble merger, and bubble collapse that we refer to as particle beating (Fig.~\ref{fig:fig1}e). The robustness of this self-sustained cycle is evidenced by the tracked coordinates of the two particles over a course of 280s (Fig.~\ref{fig:fig1}f and Supplementary Movie 2), which contrast the single particle scenario where practically no motion was observed. Notably, while the central challenge in self-oscillatory systems is to keep them away from equilibria~\cite{Hua2021,Shen2021}, such states are virtually eliminated from our system by the effectively instantaneous nature of bubble collapse.

We monitored the oscillatory behaviour of the system by tracking its breathing radius $r(t)$ over time, defined as:
\begin{equation}
    r(t) = \frac{1}{N}\sum_{i=1}^{N}\sqrt{(x_i(t)-\Bar{x})^2+(y_i(t)-\Bar{y})^2} \tag{2} 
    \label{eq:breathing_radius}
\end{equation}
for a collection of $N$ particles each with coordinate $(x_i(t),y_i(t))$ at time $t$. In other words, $r(t)$ is the Euclidean distance from the collective's centroid $(\Bar{x},\Bar{y})$ to each particle, averaged over all particles (see annotations in Fig. \ref{fig:fig1}e). The system’s periodic beating is evident in the time evolution of $r(t)$ (Fig.~\ref{fig:fig1}g, left panel), the limit cycle of its $r(t)$ phase portrait (Fig.~\ref{fig:fig1}h, Methods), as well as the narrow peak in the recurrence time histogram (Fig.~\ref{fig:fig1}i, Methods). Taken together, these analyses serve as conclusive evidence of the long-term stability of system oscillations. The analysis in Fig.~\ref{fig:fig1}i shows a period of 3.2s for the two-particle system in 10.7wt\% H$_2$O$_2$, consistent with Fig.~\ref{fig:fig1}g and Supplementary Movie 2. The period remains constant throughout as revealed by the moving-window recurrence analyses (Supplementary Fig.~6, Methods), since a negligible 0.02\% of the fuel is consumed over 280s based on stoichiometry. Furthermore, the oscillation amplitude and periodicity are shown to be resilient towards various forms of perturbations (Supplementary Fig.~9). We developed a mechanistic model based on calculated $\mathbf{F}_\text{g}$, $\mathbf{F}_\text{c}$, and the non-Stokesian drag force $\mathbf{F}_\text{d}$ (Supplementary Note 1), and found that it captured even the detailed dynamics of the breathing radius' time evolution (Fig.~\ref{fig:fig1}g right panel, also Supplementary Fig.~5). We verified the consistency of the beating frequency across 8 sets of independent experiments with 10.7wt\% H$_2$O$_2$ in Fig.~\ref{fig:fig1}j. Additionally, the beating frequency's dependence on H$_2$O$_2$ concentrations points to a mechanism for exerting fine control over the beating frequency, as predicted by our mechanistic model based on a Langmuir-Hinshelwood kinetics of the catalytic surface (Fig. \ref{fig:fig1}j)~\cite{Lin1998,Plauck2016}. In Supplementary Figs. 7 and 8, we further explored the dependence of the oscillation amplitude and frequency on H$_2$O$_2$ volume and particle size. Of note, the stable emergent self-oscillation presented in Fig.~\ref{fig:fig1} does scale down to 250$\mu$m-and 100$\mu$m-diameter particles.

%%%%%%%%%%
% Figure 2
\begin{figure*}[pt]
    \centering
    \includegraphics[width=0.69\linewidth]{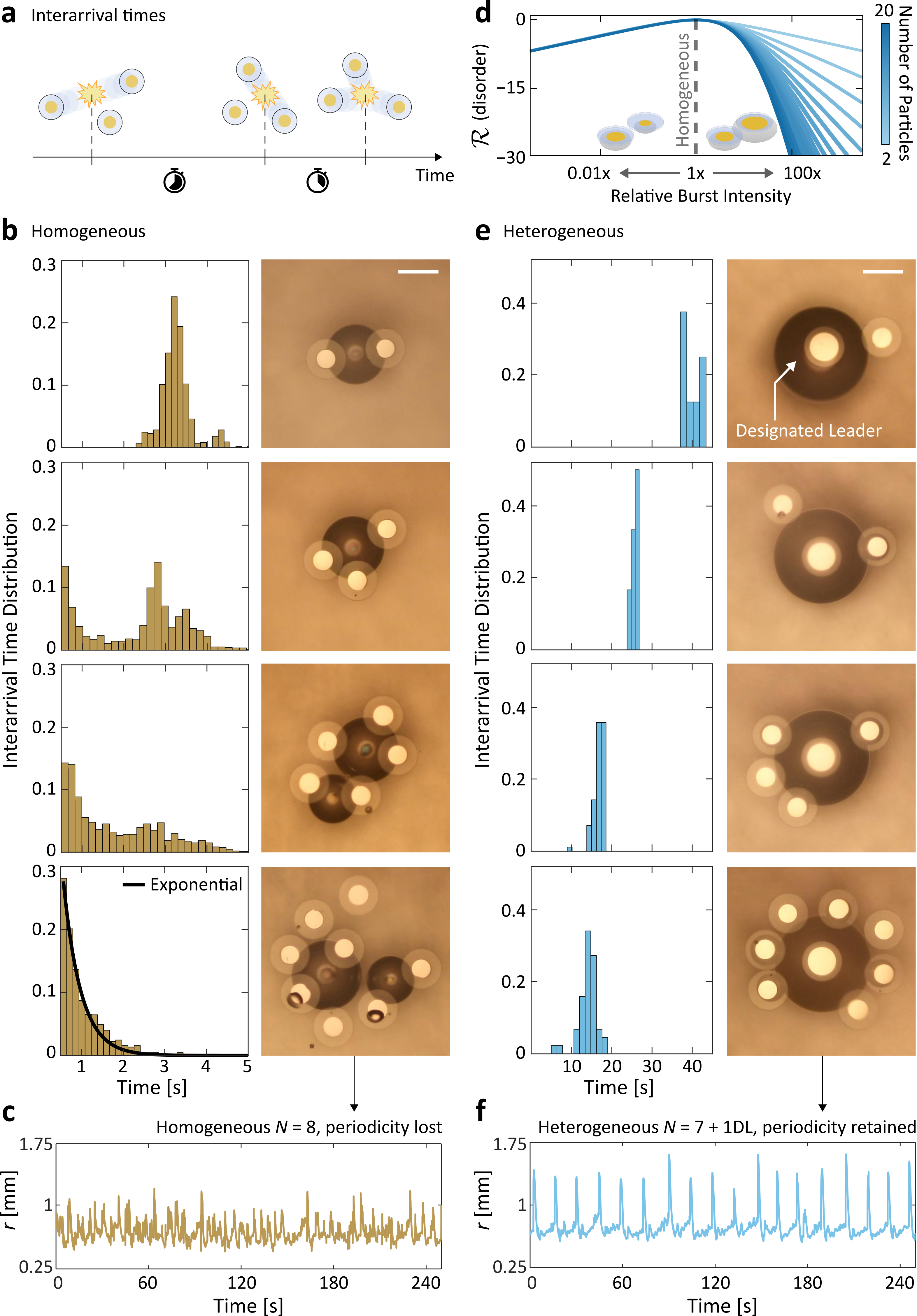}
    \caption{\small \textbf{Observations of emergent order via symmetry-breaking.} 
    \textbf{a}, Schematic of interarrival times in a system of beating microparticles, defined as the time that transpires between two consecutive bubble collapses. The interarrival time distribution should be tight (i.e., a single peak) in a perfectly periodic system, and broad in an aperiodic system.
    \textbf{b}, (top to bottom) Interarrival time distributions and optical micrographs for homogeneous systems of $N = 2, 3, 5,$ and 8 identical particles. As $N$ increases, the collective system periodicity gradually decays and transitions to an exponential interarrival distribution at $N=8$ (bottom, black curve). Scale bar, 500$\mu$m.
    \textbf{c}, Indeed, we observe that the breathing radius of a homogeneous $N = 8$ system is not periodic. 
    \textbf{d}, Asymmetry-induced order across $N$ predicted by Rattling Theory. A quantification of collective disorder, the system's Rattling $\mathcal{R}$ is predicted to be lower (i.e. more orderly) if the relative burst intensity of one particle is increased beyond or decreased below 1x, which signifies homogeneity. This is experimentally realized by modulating the Pt patch size on a ``designated leader'' (DL) particle relative to the others. The curves are offset to make all $\mathcal{R}=0$ at 1x intensity to highlight the effect of system heterogeneity on Rattling. See Supplementary Note 2 for a detailed discussion of the analytical model.
    \textbf{e}, Same as (\textbf{b}), but for heterogeneous systems of equal particle numbers, where the DL broke the permutation symmetry. In contrast to the homogeneous systems (\textbf{b}), they remain robustly periodic across $N$. It is important to recognize that the polymeric disc size of a DL is unchanged. Scale bar, 500$\mu$m.
    \textbf{f}, Breathing radius for an 8-particle DL system (i.e., $N = 7+1$DL), which reliably beats periodically. The period of 14.2s extracted from $r(t)$ coincides with the most probable interarrival time in (\textbf{e}, bottom).}
    \label{fig:fig2}
\end{figure*}

%%%%%%%%%%%
% Section 2
\subsection*{Persistent periodicity via symmetry-breaking}
\label{sec:sec2}
\begin{figure*}[!t]
    \centering
    \includegraphics[width=1.0\linewidth]{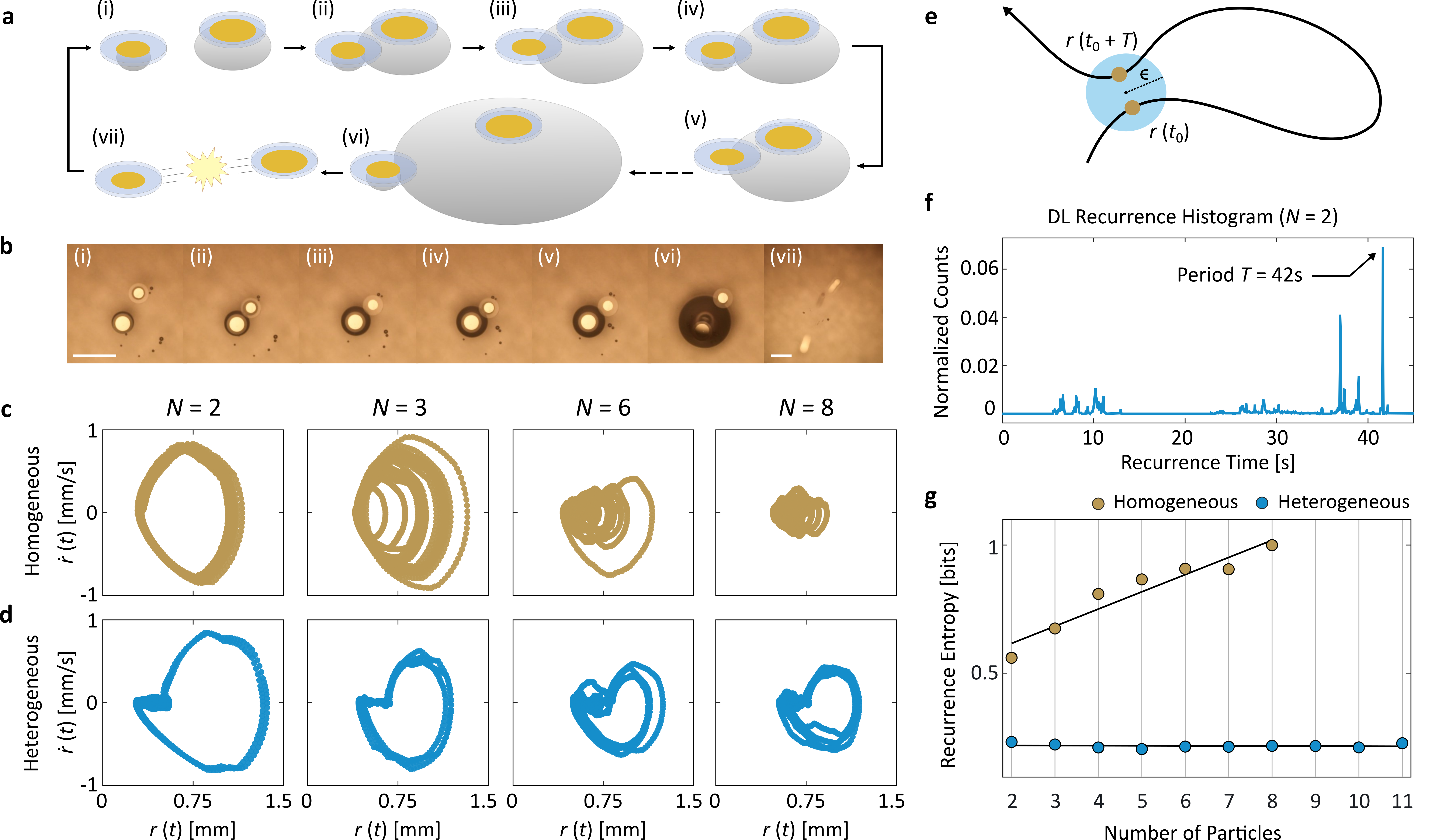}
    \caption{\small \textbf{Designated leaders induce periodic limit cycles.}
    \textbf{a},\textbf{b}, Features of DL beating explained with schematic (\textbf{a}) and micrograph sequence (\textbf{b}) of a 2-particle heterogeneous system. The leader particle is able to grow a large bubble promptly and subsume the smaller bubbles of neighbouring particles across several rounds of bubble coalescence. Scale bars, 1mm.
    \textbf{c},\textbf{d}, Phase portraits of homogeneous (\textbf{c}) and heterogeneous (\textbf{d}) systems of $N=2,3,6,$ and 8. Only the latter is able to maintain the closed-loop orbits at high particle counts.
    \textbf{e}, Schematic of recurrence time calculation. The recurrence time is the time it takes to return from a given system configuration to the neighborhood of said configuration (see Methods). 
    \textbf{f}, Recurrence histogram compiling all of the recurrence times observed across experiments of the 2-particle heterogeneous system ($N=1+1$DL). 
    \textbf{g}, Recurrence entropy as a function of $N$ for both homogeneous (yellow) and heterogeneous/DL (blue) systems. Low recurrence entropy is a quantitative indicator of periodic behaviour. The homogeneous system's recurrence entropy trends upward, suggesting a decay in periodicity, while the DL system's entropy remains low in accordance with its observed periodicity even at high $N$.}
    \label{fig:fig3}
\end{figure*}

Our findings in Figs. \ref{fig:fig2} and \ref{fig:fig3} show that the stable emergent self-oscillation can be extended well beyond $N=2$, although curiously only when the system's permutation symmetry is broken and not in a homogeneous system of identical particles. We extracted the bubble burst interarrival time statistics by tracking the time that transpires between each pair of consecutive bursts in recorded experiments (Fig.~\ref{fig:fig2}a). In homogeneous systems of identical particles (Fig.~\ref{fig:fig2}b), we show that the likelihood of periodic beating dwindles gradually with rising particle counts $N$, reflected in the progressive decay in the sharpness and amplitude of the initial 3.2s peak corresponding to periodic beating. The decay of collective periodicity is accompanied by an increase in the probability mass of frequent and unpredictable bubble bursts taking place less than a second from one another---a result of bubble mergers and collapses among subsets of particles (see representative $N=5$ and 8 micrographs in Fig.~\ref{fig:fig2}b). Interestingly, we find that the interarrival time distributions of systems beyond $N=7$ become statistically indistinguishable from those of a Poisson process (Fig.~\ref{fig:fig2}b, bottom panel)~\cite{Gallager2013}. This shows that our system’s phenomenology can remarkably vary from coordinated and reliable periodic beating to independent and effectively stochastic bubble bursts merely as a function of $N$. The breathing radius trajectory in Fig.~\ref{fig:fig2}c confirms the loss of periodicity, as no structure can be discerned from the noisy low-amplitude fluctuations.

The gradual transition towards aperiodicity in Figs.~\ref{fig:fig2}b and c points to the nominal fragility of periodic beating as the system size increases. Reasoning that the deliberate introduction of heterogeneity has been shown to produce asymmetry-induced order~\cite{Medeiros2021} in complex networked systems~\cite{Zhang2021_sync,Zhang2017_asymmetry,Nicolaou2021}, we investigated the effect of permutation symmetry-breaking on the robustness of particle beating across system sizes. Based on the role buoyancy plays in the beating physics (Supplementary Note 1), we hypothesized that particles could be made dynamically distinct from one another by controlling the relative size of their accompanying bubble. We tested the impact of this heterogeneity on collective order with Rattling Theory~\cite{Chvykov2021,Chvykov2018}. This thermodynamic theory explains the way in which correlations among driven degrees of freedom give rise to system-level fluctuations that govern the long-term stability of system configurations. The magnitude of these fluctuations, as quantified by Rattling $\mathcal{R}$, serves as an index describing the system's degree of disorder. Since lowering $\mathcal{R}$ requires substantial correlations among degrees of freedom, systems in low-$\mathcal{R}$ configurations often exhibit emergent order. 

We constructed a theoretical model that analytically connects a bubble's relative size with its contribution to system-level fluctuations, and in turn collective order (Supplementary Note 2). The model's predictions in Fig.~\ref{fig:fig2}d suggest that any deviation in a single particle's bubble size relative to the rest of the ensemble (i.e., with relative burst intensity away from 1x) results in a more orderly system as quantified by lower $\mathcal{R}$. Interestingly, the reduction in $\mathcal{R}$ is found to be particularly significant when a bubble larger (and stronger) than its peers is introduced, which we confirmed with experiments. We note that this novel mechanism for asymmetry-induced order applies to a broad class of complex systems wherein parametric heterogeneities control the fluctuations of strongly interacting elements (see Supplementary Note 2).

We broke the permutation-symmetry of the original system experimentally by adding a ``designated leader'' (DL) particle with an enlarged Pt patch of radius 175$\mu$m (Fig.~\ref{fig:fig2}e). Note that since the nanometre-scale thickness of the Pt layer is negligible compared to that of the unchanged 10$\mu$m-thick polymeric microdisc, the DL design does not alter the particle’s volumetric geometry. However, the heterogeneity among the catalytic surface areas translates directly to unequal bubble growth rates between the DL and its neighbours, which in turn drastically affects their collective dynamics in accordance with our theoretical predictions in Fig.~\ref{fig:fig2}d: We observe robustly periodic bubble collapses across $N$ in the sharp peaks of the interarrival distributions in Fig.~\ref{fig:fig2}e, suggesting that DLs are able to sustain the periodicity of particle beating even at high particle counts. Figure~\ref{fig:fig2}f depicts the time evolution of the breathing radius for a system of $N=7+1$DL particles (see also Supplementary Movie 4). In contrast to the homogeneous $N=8$ system (Fig.~\ref{fig:fig2}c), the heterogeneous DL system exhibits a stable long-term self-oscillation with a period of 14.2s, owing to the broken permutation symmetry.

%%%%%%%%%%
% Figure 3

Figures~\ref{fig:fig3}a(i-vii) and~b(i-vii) explain the microscale physics arising from the intentionally broken symmetry (see also Supplementary Movie 3). When a DL particle with an enlarged Pt patch is paired with a non-DL particle, the heterogeneity in bubble sizes leads to the subsumption of the non-DL particle bubble into the DL bubble upon contact (Figs.~\ref{fig:fig3}a(ii-v) and b(ii-v)). This coalescence behaviour is distinct from that of equal-sized bubbles previously shown in Fig.~\ref{fig:fig1}b, where an unstable merged bubble forms halfway between the particles. Instead, the merged bubble sticks to the former location of the large parent bubble underneath the DL particle, seen in Figs.~\ref{fig:fig3}a(iii) and~(v). This behaviour falls under the sticking bubble regime in the literature, a phenomenon long observed in experiments~\cite{Chen2009_microbubbles,Moreno2018} but only recently thoroughly studied and theorized in a catalytic H$_2$O$_2$ bubble system~\cite{Lv2021}. Importantly, contrary to the more intuitive moving bubble regime where the merged bubble sits at the centre of mass of its parents~\cite{Weon2012,Chen2017_microbubbles}, the coalescence behaviour transitions into the sticking regime only as the parent bubbles differ sufficiently in size~\cite{Lv2021}, or, in other words, with sufficient particle heterogeneity. As shown in the rest of Figs.~\ref{fig:fig3}a and b, the two particles in the system undergo several rounds of small-scale bubble coalescence, eventually causing the DL bubble to collapse. We find that the bubble’s rupture radius is approximately 1.7 times larger than that for a homogeneous system shown in Fig.~\ref{fig:fig1}f, stabilized by the particle sitting directly on top. This contributes to an even lower-frequency chemomechanical oscillation (Figs.~\ref{fig:fig2}f and~\ref{fig:fig3}f) than that previously observed in homogeneous systems (Fig.~\ref{fig:fig1}i and~\ref{fig:fig2}b). 

%%%%%%%%%%
% Figure 4
\begin{figure*}[!t]
    \centering
    \includegraphics[width=1.0\linewidth]{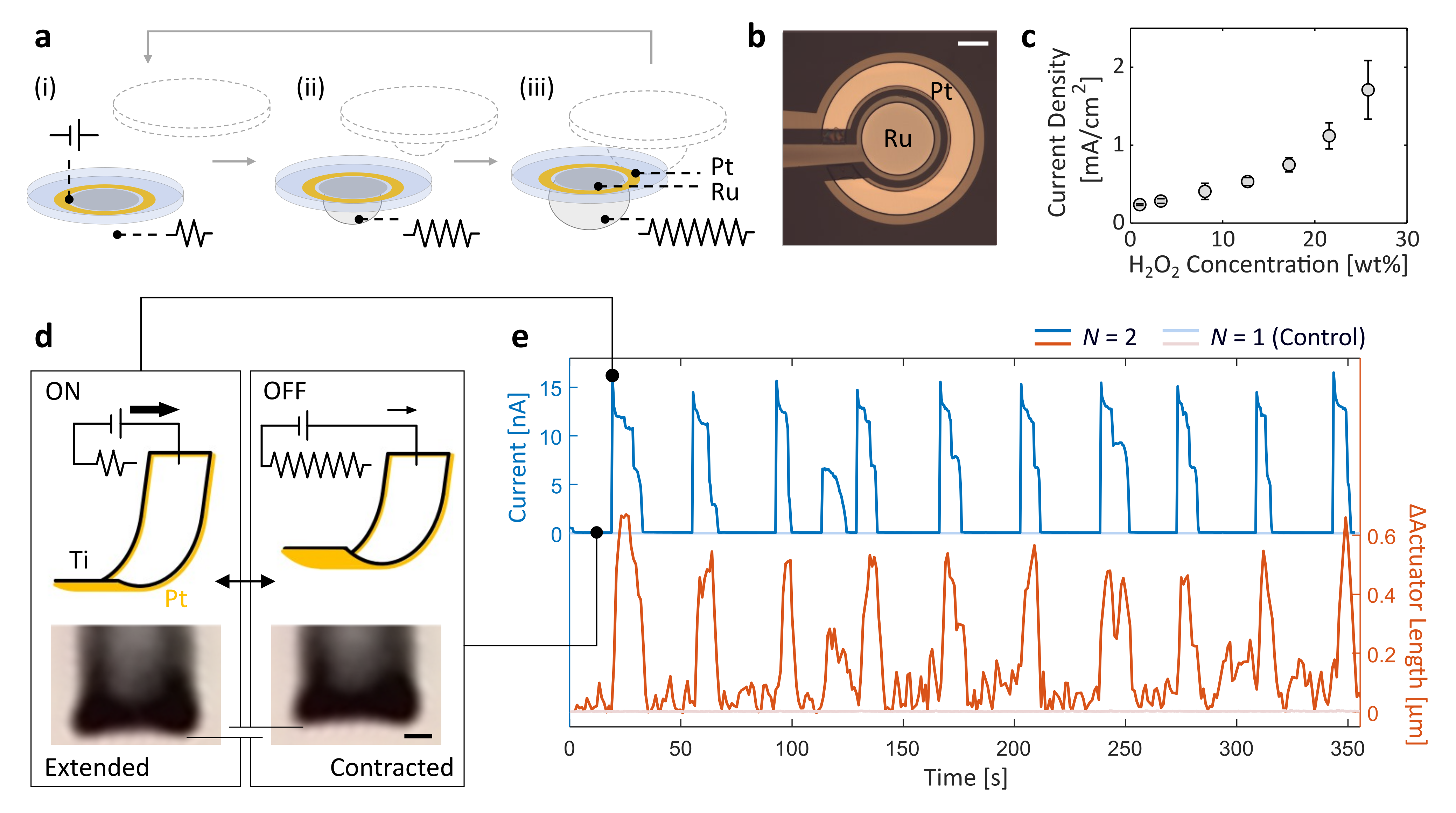}
    \caption{\small \textbf{Self-organized oscillation powers a microrobotic arm.}
    \textbf{a}, Schematics of the generation of an oscillatory electrical current from chemomechanical beating. The pair of metals (Pt-Ru or Pt-Au) patterned on a polymer base  constitute the electrodes of a H$_2$O$_2$ fuel cell, which serves as an on-board voltage source. The periodic bubble growth and collapse in a beating system separately modulates the electrical resistance between the electrodes, leading to an oscillatory current.
    \textbf{b}, Optical micrograph of a typical Pt-Ru fuel cell particle. The entire surface, less the electrode area, is passivated with a thin layer of insulating SU-8 polymer (shaded). The metallic leads on the left are not necessary for device operation and are included to facilitate measurement. Scale bar, 100$\mu$m.
    \textbf{c}, Short-circuit current density as a function of H$_2$O$_2$ concentration for a Pt-Ru device.
    \textbf{d},\textbf{e}, Cyclic motion of a microrobotic actuator driven by the oscillatory current. The schematics and micrographs in (\textbf{d}) show the extended and contracted states of the actuator respectively under the ON and OFF current conditions, as modulated by the bubble size. The current measurement over time and the actuator length change (\textbf{e}) closely match, confirming that the cyclic actuation is driven by the oscillatory current, which itself is emergent from the particle beating. Scale bar, 2$\mu$m.
    }
    \label{fig:fig4}
\end{figure*}

Figures~\ref{fig:fig3}c and~d contrast the breathing radius phase portraits between homogeneous and heterogeneous systems of different $N$. We observe that the homogeneous systems experience a decay of periodicity evidenced by the gradual collapse of limit cycle orbits in its phase portraits as a function of $N$, consistent with trends in Fig.~\ref{fig:fig2}b. In contrast, the heterogeneous systems' limit cycles are robust to variations in $N$, retaining their closed-loop phase-space orbits. To rigorously quantify the effect that DLs have on collective periodicity, we analysed the recurrence structure of the dynamical trajectories across system sizes (see Methods)~\cite{Eckmann1987}. As sketched in Fig.~\ref{fig:fig3}e, recurrence analyses capture the dynamical properties of system behaviours by measuring the time the system takes to return to a given state’s neighbourhood. The set of all such time intervals is compiled into a recurrence histogram (Fig.~\ref{fig:fig3}f) whose recurrence entropy can be used to quantify the complexity of dynamical trajectories~\cite{Marwan2007}, with perfect periodicity corresponding to zero entropy. 

The linear entropy increase for homogeneous systems as a function of $N$ (Fig.~\ref{fig:fig3}g) corresponds to the increasing disorder in the system's recurrences that is consistent with the progressive loss of periodicity observed in Figs.~\ref{fig:fig2}c and~\ref{fig:fig3}c. Also in accordance with earlier qualitative trends in Figs.~\ref{fig:fig2}f and~\ref{fig:fig3}d, the recurrence entropy of the DL system is locally invariant to changes in $N$, thereby providing quantitative evidence of the robustness of the periodic beating induced via symmetry-breaking. While we find that the system's invariance to particle number holds up to $N=11$, we leave the study of larger particle systems for future work (Supplementary Figs.~10 and~11).

%%%%%%%%%%%
% Section 3
\subsection*{Self-oscillating microgenerators}
\label{sec:sec3}
Through a simple modification to the particle design, we are able to harness the robust chemomechanical beating to generate an oscillatory electric signal. As illustrated in Fig.~\ref{fig:fig4}a and b, we fabricated particles with a Pt pattern closely lined up with (though spatially separate from) an additional metal patch of either Au or Ru (see Methods). With the bimetallic design, the previously auto-redox catalytic decomposition of H$_2$O$_2$ on Pt is in part separated into an oxidation half-reaction on Pt and a reduction half-reaction on Ru (Au)~\cite{Wang2006_nanomotor,Paxton2006,Wang2013_efficiency}:
\begin{chemmath}
  \text{Pt:}\ \ H_2O_2 \reactrarrow{0pt}{0.5cm}{}{} O_2+2H^+ + 2e^-\notag\\
\end{chemmath}
\vspace{-0.2in}
\begin{chemmath}
  \text{Ru (Au):}\ \ H_2O_2 +2H^+ + 2e^-\reactrarrow{0pt}{0.5cm}{}{} 2H_2O\tag{3}\label{eq:elec_rx}\\
\end{chemmath}
\vspace{-0.2in}
\begin{chemmath}
  \text{Overall:}\ \ 2H_2O_2 \reactrarrow{0pt}{0.5cm}{}{} 2H_2O + O_2\notag.\\
  \setcounter{equation}{4}
\end{chemmath}

\noindent Consequently, a potential difference is established at the two electrodes that essentially transforms the particle into an on-board fuel cell.
These same principles have been previously used to generate voltages in nanomotors, where bimetallic rods and nanoparticles are propelled electrokinetically by the accompanying electric field~\cite{Wang2013_PNAS,Lee2014_nanomotors,Zhang2021_NRC}. A micrograph of our fabricated prototype is displayed in Fig.~\ref{fig:fig4}b. Note that the metallic leads extending outwards were added to facilitate electrical characterization of the devices and are not necessary to their operation. The leads were passivated and hence do not participate in any electrochemical reactions. The Pt-Ru and Pt-Au fuel cell devices measured open-circuit voltages of 144.9mV $\pm$ 2.4 and 21.4mV $\pm$ 3.5, respectively, in a 25.8wt\% H$_2$O$_2$ solution with 0.075M KNO$_3$ added for conductivity (see Methods and Supplementary Fig.~13). These values are in line with prior mechanistic studies~\cite{Wang2006_nanomotor,Paxton2006} (Supplementary Note 4). Under the same conditions, the Pt-Ru fuel cell delivers a short-circuit current density of 1.71mA/cm$^2\pm$ 0.38 and a current of 56.7nA $\pm$ 12.4. As a benchmark, a significantly larger 1.5 $\times$ 6cm thermo-mechano-electrical self-oscillator reported recently recorded a peak current of $\sim$47nA~\cite{Wang2018_selfpropulsion}. The dependence of the current density on H$_2$O$_2$ concentration is summarized in Fig.~\ref{fig:fig4}c (also Supplementary Fig.~14).

As before, the system's collective beating drives the synchronized formation and collapse of bubbles on each particle. However, unlike previous experiments, here the instantaneous size of the bubble also modulates the electrical conductance from one electrode to the other (Fig.~\ref{fig:fig4}a, $N=2$ for demonstration). This effect, in conjunction with the fuel cell’s voltage, enables the onboard generation of oscillatory currents that are in phase with the mechanical beating (Supplementary Fig. 16). In a Pt-Ru device, we observe that the ON/OFF ratio between maximal and minimal currents can exceed $10^6$, corresponding to when the bubble is absent and at its threshold size. Importantly, the same chemical energy harnessed from the environment is used to simultaneously drive the mechanical oscillation, generate the electrical voltage, and modulate the electrical conductance. Multifunctionality of this kind is emblematic of emerging paradigms such as embodied energy~\cite{Aubin2022}, and is crucial to the development of efficient microsystems. 

Figures~\ref{fig:fig4}d and e exemplify the beating system’s capability to cyclically drive a microrobotic load with its self-generated oscillatory electrical current. In this proof-of-concept demonstration, we wired the Ru electrode of a fuel cell particle to a state-of-the-art Pt-Ti electrochemical microactuator (see Fig.~\ref{fig:fig4}d and Methods), originally invented for a tethered sub-100$\mu$m walker~\cite{Miskin2020}. In our experimental configuration, charged species from the electrolyte is desorbed from the Pt surface of the bimorph microactuator as current passes through, causing it to deswell and its curvature/length to change. Evident in Fig.~\ref{fig:fig4}e, the periodic actuation of the bimorph (red curve, representative snapshots in Fig.~\ref{fig:fig4}d, also Supplementary Movie 5) is driven by the periodic spikes in the current signal (blue curve), which in turn is modulated by the chemomechanical beating of two particles. Because the outer radius of the Pt electrode (Fig.~\ref{fig:fig4}b) exceeds the 125$\mu$m patch radius of a standard particle, the system is stabilized by the added heterogeneity, which also explains the observed sub-0.03Hz beating frequency. In contrast, the control experiments in Fig.~\ref{fig:fig4}e show the actuator idling in the absence of a second particle and hence any mechanical beating. By harnessing the emergent power generation of an ensemble of microparticles, we have demonstrated the design and modular interoperability of key microrobotic components---energy sources and locomotive elements---based on the physics of self-organization.

%%%%%%%%%%%%
% Discussion
\section*{Discussion}
\label{sec:sec4}
Through the discovery of physical mechanisms for asymmetry-induced order, we constructed self-oscillating electrical generators capable of powering on-board microrobotic components from the interactions of simple microparticles. Our results stand in contrast to more traditional microrobotic approaches focusing on the design of intricate electromechanical assemblies to produce alternating electrical currents~\cite{Akbar2021}. By relying on our system's self-organized behaviours, we circumvented the design of complex contraptions to harvest and transduce chemical energy into periodic electrical and mechanical work---a crucial step towards fully-autonomous microrobots~\cite{Brooks2020, Aubin2022}. The use of on-board electrical currents will enable the integration of sensors and computational elements to enrich physical microparticle interactions~\cite{Yang2021_memristor}, forming the basis for future collectives wherein the long-envisioned potential of complex inter-particle communications can be implemented~\cite{Solovev2013_collective}. We plan on extending our approach into studying larger collections of microparticles in search of general principles for the top-down design of active matter systems, where an understanding of system symmetries and environmental forcing may enhance their task-capability. Unifying perspectives from their respective fields, our work suggests that future microrobots and active matter systems may become more robust and task-capable when we design them to exploit the physics of the environments they inhabit.

%%%%%%%%%
% Methods
\section*{Methods}

\subsection*{Fabrication and liftoff of microparticles}
The fabrication process is summarized in Supplementary Fig.~4. SU-8 2010 photoresist was spun on a Si wafer at 3000rpm for 1 minute. It was baked at 65$^\circ$C for 1 minute and 95$^\circ$C for 2 minutes. The SU-8 discs were defined by exposure with a Karl S\"{u}ss MA6 Mask Aligner at a dose of 140mJ (365nm). The wafer was baked post-exposure at 65$^\circ$C and 95$^\circ$C respectively for 1 and 2.5 minutes. The resist was developed in SU-8 developer for 2.5 minutes, soaked in isopropanol, and blow dried. The wafer was optionally hard baked at 115 to 180$^\circ$C for 10 minutes to 2 hours.

LOR 3A photoresist was spun onto the sample at 1000rpm for 1 minute. This was optionally followed by a second spinning step at 2000rpm for 30 seconds to ensure that the coating was uniform at the periphery. The sample was baked at 180$^\circ$C for 4 minutes. Shipley S1818 photoresist was spun at 2000rpm for 1 minute and subsequently heated at 115$^\circ$C for 1 minute. The LOR and Shipley parameters were optimized to ensure a full coverage over the 10$\mu$m-thick SU-8 discs. The sample was aligned and exposed at 140mJ (405nm). It was then developed in AZ 726 MIF for 1.5 minutes. The sample was washed with running DI water and blow dried. 

The Pt metal patches were deposited with a Denton e-Beam Evaporator. A typical patch consists of 5nm of Cr or Ti adhesion layer and 50nm of Pt. The photoresists were stripped in Remover PG. 

The fabricated microparticles were lifted off the wafer substrate in 45$^\circ$C 1M KOH solution, which etched away Si (Supplementary Fig.~4b). The process typically took 30 to 50 minutes. The microparticles were collected by a transfer pipette and then washed repeatedly with DI water until the solution’s pH was neutral. Alternatively, the microparticles were first coated with PMMA A4 (polymethyl methacrylate) before being lifted off in 90$^\circ$C 1M KOH solution (Supplementary Fig.~4c). The microparticle array on the PMMA sheet was picked up by a clean piece of wafer. The PMMA was carefully dissolved away with acetone and the particles were washed by and stored in DI water.

\subsection*{Experimental characterization of beating behaviour}
In a typical experiment, 1mL of H$_2$O$_2$ solution (10.7\% unless otherwise noted, VWR International, LLC, Radnor, PA) is dispensed gently onto a polystyrene Petri dish (VWR International, LLC, Radnor, PA). Two methods were used to transfer the micro-oscillators from their vial to the H$_2$O$_2$ droplet. In the ``wet'' method, they could be collected with a narrow-tipped transfer pipette along with a small amount of water, and subsequently transferred onto the droplet. The introduction of a minor amount of diluent as well as the occasional need to flip over a particle can be avoided with an alternative ``dry'' process. First, a particle was wet transferred onto a glass slide with a transfer pipette. Excess water was carefully wiped off while the particle was not completely dried. A drop of H$_2$O$_2$ solution was then added. This step allowed the operator to check the orientation of the particle on the glass slide prior to its transfer to the droplet. A quartz NMR sample tube was used to directly pick up the particle dry, a process assisted by surface tension. Note that the other end of the tube was, of course, capped. Lastly, the dry particle with the correct orientation was gently placed atop the 1mL H$_2$O$_2$ droplet under the camera.

The beating behaviour was recorded as 30fps videos with a Canon Rebel T6i camera (Canon U.S.A., Inc., Huntington, NY). The optical system comprised a magnification lens (MVL12X20L), a coaxially focusable zoom lens (MVL12X3Z), and an extension tube (MVL12X3Z), all purchased from Thorlabs, Inc., Newton, NJ. The setup followed that described previously in~\cite{Zhao2020_2Dmaterials}. The illumination source was a MI-150 Fiber Optic Illuminator from Edmund Optics Inc., Barrington, NJ.

\subsection*{Phase and recurrence analyses of particle beating}
The recorded videos of the beating systems were processed with the Image Processing Toolbox of MATLAB (MathWorks, Inc., Natick, MA). The particle centres were identified from each frame of the videos with the standard \texttt{imfindcircles} function, a circle-finding algorithm based on circular Hough transform~\cite{MatlabCircleFind}. Given a collection of particle trajectories from an experimental trial, the main observable from which to construct the phase portraits shown in Fig.~\ref{fig:fig3} was the breathing radius $r(t)$ as defined in Eq.~(\ref{eq:breathing_radius}). The phase portraits were then constructed by plotting the coordinates of $v(t)=[\dot{r}(t),r(t)]$ after applying a low-pass filter, and the time-derivative of the breathing radius was estimated via finite differencing.

Equipped with the dynamical observables defined above, the recurrence properties of a system can be analysed by finding how often and how quickly the system returns to a neighborhood of $v(t)$. Hence, for a given experiment comprised of $K$ samples we collect data at times $t_i = i\Delta t,\ \forall i \in \{0,\cdots,K-1\}$ with sampling rate $\Delta t$. While in principle this is all one needs in order to quantify recurrence statistics~\cite{Eckmann1987}, an additional step must be taken in order make the calculation robust. We augmented our $v(t_i)$ vectors by ``embedding'' the time-series according to an integer parameter $m$~\cite{Marwan2007}. This resulted in a modified set of coordinates, $v_m(t_i) = [v(t_i), \cdots, v(t_{i+m-1})]^T$, from which to robustly calculate our recurrence statistics. Finally, to derive the recurrence properties of a system from an experimental dataset we calculated its recurrence set
\begin{equation}
    R_s=\{|t_i-t_j|:\ ||v_m(t_i)-v_m(t_j)||<\epsilon, \forall i,j\},\tag{4}
    \label{eq:recurrence_set}
\end{equation}
over all valid indices. Note that $m$ and $\epsilon$ are a fixed choice of positive non-zero embedding dimension and neighborhood size parameters, respectively. With this set now defined, we could calculate a recurrence histogram from the set $R_s$ using any standard scientific computing package, as in Fig.~\ref{fig:fig3}. Additionally, we note that the histogram can be normalized into a pseudo-probability distribution that expresses the likelihood $p(T)$ that a system exhibits a recurrence after $T$ seconds. The dominant frequencies plotted in Figs.~\ref{fig:fig1} and~\ref{fig:fig3} and Supplementary Fig.~6 were computed from the $T$ of maximum likelihood from the corresponding recurrence analyses.

As we are interested in characterizing the onset of periodicity across collectives of beating particles, we must construct a measure capable of differentiating the diversity of behaviours we observed. For this purpose, we made use of the entropy of the recurrence probability distributions. As an example, consider a system with a single perfectly oscillatory mode. Then, its recurrence distribution would be a delta function corresponding to its period of oscillation, and thus have zero entropy. If one were to introduce noise or uncertainty into that single oscillatory mode, then probability mass would spread around the delta peak and generate non-zero entropy. Likewise, if the system were to have multi-modal (but deterministic) oscillation, probability mass is now shared between the peaks of the distribution, leading to non-zero entropy. 

As the behaviour of a system becomes increasingly complex, it has been shown that the recurrence distribution entropy is a useful metric to quantify this shift that has known connections to both Kolmogorov-Sinai and R\'{e}nyi entropies~\cite{Faure1998}, as well as the correlation sum in chaos theory~\cite{March2005}. However, in order to compare the recurrence entropies of systems with different maginitude- and time-scales, we first normalized our data in two ways. First, we applied min-max normalization to the coordinates of $p(T)$, which allows one to use the same $\epsilon$ in the calculation of the recurrence set. Second, we normalized the elements of $R_s$ according to its maximum (while keeping the number of histogram bins constant across systems) in order to study the structure of system recurrences without confounding variables. The result of this process can be seen in Fig.~\ref{fig:fig3}.

\subsection*{Fuel cell fabrication}
LOR 20B photoresist was spun onto a Schott Borofloat 33 wafer (UniversityWafer, Inc., Boston, MA) at 3000rpm for 1 minute and baked at 180$^\circ$C for 4 minutes. Shipley S1805 photoresist was spun at 3000rpm for 1 minute and baked at 115$^\circ$C for 1 minute. The sample was exposed at 82.5mJ (405nm). It was then developed in Microposit MF-319 developer for 65 seconds. The sample was washed with running DI water and blow dried. A Denton e-Beam Evaporator was used to deposit 10nm of Ti and 50 to 100nm of Pt. The photoresists were stripped in Remover PG. For the deposition of a second metal, be it Au or Ru, LOR and Shipley resists were spun, baked, exposed, and developed the same as described above. 10nm of Ti and 100nm of Au was deposited with an electron beam evaporator. Alternatively, 50nm of Ru was deposited as the deposition was slow. The photoresists were stripped in Remover PG. 

The SU-8, LOR, and Shipley resists were all purchased from Kayaku Advanced Materials, Inc., Westborough, MA, in addition to the SU-8 developer, MF-319 developer, Remover PG, and PMMA. The AZ 726 MIF was purchased from MicroChemicals GmbH, Ulm, Germany.

Finally, a passivation layer of SU-8 was defined on top of the metal electrodes. For the convenience of the electrical measurements that followed, the SU-8 were patterned as either 5mm-by-5mm or 11mm-by-11mm square islands with the active electrode area at the centre exposed. SU-8 2002 was used but the precise thickness was inconsequential.

\subsection*{Fabrication and characterization of microactuators}
The Pt-Ti bimorph microactuators were fabricated on a Cu sacrificial layer at University of Pennsylvania’s microfabrication facility according to the procedures previously reported~\cite{Miskin2020}. The actuators were lifted off overnight in a 4mg/mL ammonium sulfate solution, which etched away the Cu substrate. The actuators were subsequently transferred to a phosphate-buffered saline (PBS) solution. 

In Fig.~\ref{fig:fig4} of the main text, the bimorph microactuators were cyclically driven by the oscillatory electrical current signal generated by the oscillatory beating between a Pt-Ru fuel cell device and a Pt-decorated beating particle. Each microactuator was picked up by a parylene-coated Pt-Ir monopolar electrode (PI2003X.XA3, 0.1M$\Omega$, Microprobes for Life Science, Gaithersburg, MD) in PBS. The parylene coating prevented unnecessary current leakage into the electrolyte. The Pt-Ir electrode connected to the Ru electrode of the fuel cell device via a probe station (Advanced Research Systems, Macungie, PA) and a W probe (The Micromanipulator Company, Carson City, NV). The probe station read out the real-time current with a custom MATLAB code. The Pt electrode of the fuel cell, via a W probe, was connected to a Pt wire partially immersed in the PBS solution. 

A 30\% H$_2$O$_2$ solution and a 0.5M KNO$_3$ solution were mixed at a volumetric ratio of 85:15. The salt was included to enhance the electrolyte’s electrical conductivity. For the self-oscillation to take place, 8.5$\mu$L of the prepared mixture was dropped atop a fuel cell device on the wafer. A beating particle was subsequently transferred to the same solution using the transfer method described earlier. The actuation was recorded with the same optical setup described above mounted over the probe station.

\subsection*{Actuation analysis of microactuators}
The extent of actuation as a function of time was extracted from the recorded videos described in the previous section. A standard canny edge detection algorithm with pixel magnitude thresholds was applied via OpenCV~\cite{opencv_library}. A boundary representing the outline of the actuator was extracted, which could then be used to define a coordinate system aligned and centred along the long edge of the actuator over the duration of the video---a crucial step for reducing measurement drift. From this coordinate system, the length of the actuator was then simply defined according to the nearest actuator boundary pixels along the vertical axis. Finally, in order to mitigate the effect of fluctuations and mechanical vibrations, a standard low-pass finite impulse response (FIR) interpolation scheme was applied to the actuator length signal over time~\cite{Schafer1973}.

\section*{Data availability}
\label{sec:data_statement}
The data supporting the findings of this study are available from the corresponding author upon reasonable request. 

\section*{Code availability}
\label{sec:code_statement}
The code supporting the findings of this study are available from the corresponding author upon reasonable request. 

%%%%%%%%%%%%
% References

\section*{Acknowledgements}
\label{sec:acknowledgements}
The authors are appreciative of funding from the US Army Research Office MURI grant on Formal Foundations of Algorithmic Matter and Emergent Computation (W911NF-19-1-0233) for the computational and metrological tools as well as the analysis of emergent behavior used in this work. Funding from the US Department of Energy (DOE), Office of Science, Basic Energy Sciences (grant DE-FG02-08ER46488) supported device and experimental design, fabrication and component engineering. We acknowledge helpful discussions with Dana Randall, Andrea Richa, Daniel Goldman, Jeremy England, Ana Pervan, Annalisa Taylor, Shengkai Li, Hridesh Kedia, Mahesh Kumar, Matthias K\"{u}hne, Joy Zeng, Jorg Scholvin, and Kaihao Zhang. Microfabrication for this work was performed at the Harvard University Center for Nanoscale Systems (CNS), a member of the National Nanotechnology Coordinated Infrastructure Network (NNCI), which is supported by the National Science Foundation under NSF award No. ECCS-2025158; the MIT.nano microfabrication facility at Massachusetts Institute of Technology; and University of Pennsylvania. G. Z. acknowledges the support from MathWorks Engineering Fellowship.

\section*{Author contributions} 
\label{sec:author_contributions}
J.F.Y., A.T.L., and M.S.S. conceived the experiments. J.F.Y., A.T.L., G.Z., and S.Y. fabricated the microparticles. J.F.Y. and A.T.L. carried out the collective beating experiments. J.F.Y., T.A.B., A.T.L., A.M.B., and L.N.L. processed the experimental data. T.A.B. performed analyses on the phase portraits, interarrival distributions, and recurrence histograms. T.A.B. derived theoretical results on asymmetry-induced order and Rattling model. P.C. contributed to theoretical analyses. J.F.Y. and A.T.L. studied the physics of the beating mechanism. J.F.Y. constructed the mechanical model and performed the simulations. J.F.Y., A.M.B., and M.S.S. designed the electrochemical fuel cells. J.F.Y. fabricated the fuel cells and performed the actuator experiments with G.Z. D.G.-M. and M.Z.M. fabricated the micro-actuators. J.F.Y, T.A.B., T.D.M., and M.S.S. wrote the manuscript with all authors contributing.

\section*{Competing interests} 
\label{sec:competing_interests}
The authors declare no competing interests.

\end{document}

\noindent\makebox[\linewidth]{\rule{\linewidth}{1.5pt}}

\vspace{0.0in}
    \begin{center}
        %{\Large \textbf{Emergent microrobotic oscillators via asymmetry-induced order}}\\ \\
        %\vspace{0.1in}
        {\Large Supplementary Information}
    \end{center}
	
	\vspace{-0.1in}
	
	\noindent\makebox[\linewidth]{\rule{\linewidth}{1.5pt}}
	
	\vspace{0.15in}

    \noindent\textbf{Jing~Fan~Yang}$^{*1}$ \ \ \ 
    \textbf{Thomas~A.~Berrueta}$^{*2}$ \ \ \ 
    \textbf{Allan~M.~Brooks}$^1$ \ \ \
    \textbf{Albert~Tianxiang~Liu}$^{1,3}$ \ \ \ 
    \textbf{Ge~Zhang}$^1$ \ \ \
    \textbf{David~Gonzalez-Medrano}$^4$ \ \ \
    \textbf{Sungyun~Yang}$^1$ \ \ \
    \textbf{Volodymyr~B.~Koman}$^1$ \ \ \
    \textbf{Pavel~Chvykov}$^5$ \ \ \
    \textbf{Lexy~N.~LeMar}$^1$ \ \ \
    \textbf{Marc~Z.~Miskin}$^4$ \ \ \
    \textbf{Todd~D.~Murphey}$^2$ \ \ \
    \textbf{Michael~S.~Strano}$^{\dagger1}$

    \vspace{0.05in}
    
    {\small \noindent$^1$Department of Chemical Engineering, Massachusetts Institute of Technology, Cambridge, MA, USA.\\
    $^2$Center for Robotics and Biosystems, Northwestern University, Evanston, IL, USA.\\
    $^3$Department of Chemical Engineering, University of Michigan, Ann Arbor, MI, USA.\\
    $^4$Department of Electrical and Systems Engineering, University of Pennsylvania, Philadelphia, PA, USA.\\
    $^5$Physics of Living Systems, Massachusetts Institute of Technology, Cambridge, MA, USA.}

    \vspace{0.05in}

\footnotetext{$^*$These authors contributed equally to the work.}
\footnotetext{$^{\dagger}$Corresponding author: strano@mit.edu}

\vspace{0.5in}

\tableofcontents

\clearpage
\newpage 

\addcontentsline{toc}{section}{Supplementary Notes}

\section*{Supplementary Notes}
\label{sec:notes_SI}

\subsection{Mechanistic Model and Simulation of Particle Beating}
\label{sec:mech_model_SI}
As the microparticles beat at the curved liquid-air interface of a drop of aqueous H$_2$O$_2$ solution, we start by solving the Laplace equation of capillarity~\cite{DelRio1997,Hauser2018}. We solved the following system of ordinary differential equations (ODEs) with MATLAB’s ode45 Runge-Kutta solver (MathWorks, Inc., Natick, MA):
\begin{align}
    \begin{split}
        \frac{d\rho}{ds}&=\cos \theta
        \label{eq:profile1_SI}
    \end{split}\\
    \begin{split}
        \frac{dz}{ds}&=\sin \theta 
        \label{eq:profile2_SI}
    \end{split}\\
    \begin{split}
        \frac{d\theta }{ds}&=\left\{ \begin{aligned}
          & \beta \text{, }s=0 \\ 
         & 2\beta +{{\gamma }_{\text{c}}}z-\frac{\sin \theta }{\rho}\text{, }s>0 \\ 
        \end{aligned} \right.
    \end{split}\\
    \begin{split}
        \frac{dV}{ds}&=\pi {{\rho}^{2}}\sin \theta 
    \end{split}\\
    \begin{split}
        \rho(0)&=z(0)=\theta (0)=V(0)=0
    \end{split}
\end{align}
where $\beta$ is the curvature at the apex $s = 0$ and $V$ is the volume. $\gamma_{\text{c}} = g(\rho_\text{l}-\rho_\text{a})/\gamma$ denotes the capillary constant, where $\gamma$ is the interfacial tension, $\rho_\text{l}$ density of the liquid phase, and $\rho_{\text{a}}$ that of air. $\theta$, $r$, $z$, and $s$ are defined in Supplementary Fig.~\ref{fig:drop_SI}. For each value of $\beta$, a solution to the initial value problem can be obtained which describes the profile of a Laplacian axisymmetric interface. A unique $\beta$ can be identified such that $V=V_{\text{drop}}$ and $\theta = \theta_{\text{c}}$ at the three-phase contact line. In our experimental system, the contact angle $\theta_{\text{c}} = 87.4^{\circ} $ for the peroxide-polystyrene interface in air. Supplementary Fig.~\ref{fig:drop_SI} below presents the interface profile for a series of $V_{\text{drop}}$ values. $\beta$ is solved to be 17.56m$^{-1}$ for $V_{\text{drop}}=1$mL.

\begin{figure*}[h]
    \centering
    \includegraphics[width=0.8\linewidth]{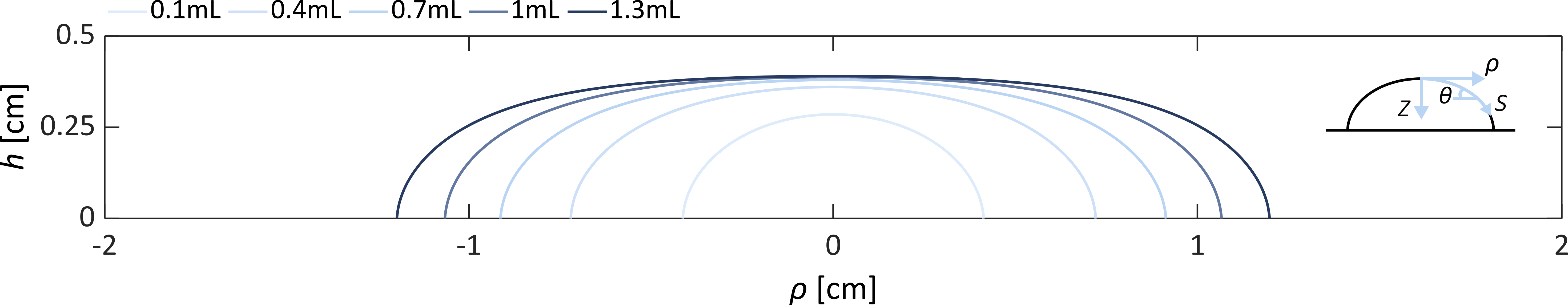}
    \caption{Coordinate system of the H$_2$O$_2$ drop and the solved interface profiles for a series of drop volumes $V_{\text{drop}}$.}
    \label{fig:drop_SI}
\end{figure*}

As discussed in the main text, the microparticles are driven outward by the collapse of a shared bubble and come together via a global and a local restorational force. As the SU-8 polymer is barely denser than the peroxide solution, the buoyancy from a small gas bubble underneath the disc is able to overcome the particle’s weight and create a net force upward. As the microparticle is constrained to the liquid-air interface, it climbs the global drop profile defined by the above solution of the Laplace equation. One can formulate the energy as the product of the particle’s vertical displacement and its weight after the subtraction of the Archimedes force. Thus, the lateral component of this global restorational force is given by:
\begin{equation}
    {{F}_{\text{g}}}=\left[ m-{{\rho }_{\text{l}}}\Lambda_{\text{b}}{{V}_{\text{b}}}(t) \right]g\frac{dz}{d\rho}
    \label{eq:Fg_SI}
\end{equation}
where $V_{\text{b}}(t)$ is the instantaneous bubble volume. Only the mass of the particle, $m$, is considered as that of the bubble is insignificant. The dimensionless factor $\Lambda_{\text{b}}$ is the volume fraction of the gas bubble lying below the undisturbed interface, as it is the displaced liquid in this region that gives rise to buoyancy. Note that we only included $\Lambda_{\text{b}}$ for the generality of Supplementary Equation~(\ref{eq:Fg_SI}). We use a $\Lambda_{\text{b}}$ of unity in the simulations hereafter in accordance with experimental observations. Given the drop profile, the force always points towards the apex. 

To quantify the ``Cheerios effect'', i.e. the inter-particle capillary attraction as a result of the local interfacial distortion, we adopt the Nicolson approximation~\cite{Nicolson1949} which assumes that (i) the horizontal force from capillary pressure is insignificant compared to that from buoyancy, and (ii) the small interfacial distortions may be superposed~\cite{Vella2005}. Prior results show that the Nicolson approximation is justified for small Bond numbers $B = R^2/L_{\text{c}}^2$, or equivalently if the floating object’s radius $R \ll L_{\text{c}} = \sqrt{\gamma/\rho_{\text{l}}g}$, the capillary length. Indeed, the $L_{\text{c}}$ of our experimental system is approximately 2.7mm, far exceeding the spatial scale of the beating physics. The surface height in the neighbourhood of a floating bubble follows:
\begin{equation}
    h(l) = -B\Sigma RK_0(l/L_{\text{c}})
    \label{eq:hcheerio_SI}
\end{equation}
where $l$ is the lateral distance from the bubble centre, $K_n$ the modified Bessel function of the second kind of order $n$, and $\Sigma$ the buoyancy-corrected dimensionless weight defined by $2\pi \gamma R B \Sigma = \left[ m-{{\rho }_{\text{l}}}\Lambda_{\text{b}}{{V}_{\text{b}}}(t) \right]g$. Supplementary Equation~(\ref{eq:hcheerio_SI}) is a simplified asymptotic result true for $l\ll L_{\text{c}}$. The lateral capillary force experienced by a bubble of volume $V_{\text{b}}'$ at a distance $l$ away is therefore:
\begin{equation}
    F_{\text{c}} = \left[ m-{{\rho }_{\text{l}}}\Lambda_{\text{b}}{V'_{\text{b}}} \right]gB^{3/2}\Sigma K_1(l/L_{\text{c}})
    \label{eq:Fc_SI}
\end{equation}
Needless to say, the capillary attraction force points towards the centre of the other particle. The $K_1(l/L_{\text{c}})$ dependence is in agreement with the results derived from an energy approach~\cite{Kralchevsky2000}. Readers are directed to \cite{Dalbe2011} for the treatment of scenarios with more than 2 particles.

We next consider the hydrodynamic interactions. In the regime of low Reynolds number and low capillary number such as our system, the drag force for an object at the liquid-air interface is expressed as:
\begin{equation}
    {{\mathbf{F}}_{\text{d}}}=-6\pi \mu {{\Lambda }_{\text{d}}}{{R}_{\text{b}}}(t)\mathbf{v}(t)
    \label{eq:stokes_SI}
\end{equation}
where $\mu$ is the liquid’s dynamic viscosity, $R_{\text{b}}$ the bubble radius, and $\mathbf{v}$ the instantaneous velocity. The drag coefficient $\Lambda_{\text{d}}$ is a scaling factor depending on the object’s geometry, its depth of immersion, the contact angle, surface tension, and the densities of the object and the liquid~\cite{Dalbe2011}. As $\Lambda_{\text{d}}$ is difficult to estimate analytically, we assume it is a constant for simplicity and leave it as one of the two free parameters we estimate from experiments, a practice consistent with published models of microparticle motion along a curved interface~\cite{Hauser2018}. 

An important additional consideration is the significantly increased drag when multiple particles approach one another, caused by the increased resistance to removing the liquid between them~\cite{Das2021}. We note this inter-particle hydrodynamic interaction particularly because of the noticeable deceleration in our beating system when the edge-to-edge distance between particles were less than $2R_{\text{p}}$ (see, for example, Fig. 1g between 68 and 69s). The approach velocity was virtually 0 right before contact, suggesting a drag significantly larger than that given by Supplementary Equation~(\ref{eq:stokes_SI}). Indeed, some previous studies predicted two floating microparticles to accelerate towards each other all the way until they collide if the Stokes’ drag expression was not corrected for inter-particle interactions~\cite{Vella2005}. 

The most numerically convenient means of accounting for said interactions is to adopt the concept of hydrodynamic mobility~\cite{Batchelor1976}, as with a number of previous works~\cite{Dalbe2011,Vassileva2005}. This correction factor as a function of the inter-particle spacing, $l$, is given by:
\begin{equation}
    G(\lambda)=1-\frac{1}{3}{{\lambda}^{-1}}+{{\lambda}^{-3}}-\frac{15}{4}{{\lambda}^{-4}}-\frac{4.46}{1000}{{\left( \lambda-1.7 \right)}^{-2.867}}
    \label{hymo_SI}
\end{equation}
where $\lambda = l/\max[R_{\text{b}}(t), R_{\text{p}}]$. $G$, which is typically multiplied to the terminal velocity, approaches 1 for large separations $\lambda\xrightarrow{}\infty$ and 0 for $\lambda=2$ when the objects contact. Equivalently, we divided the drag expression in Supplementary Equation~(\ref{eq:stokes_SI}) with $G$ in our numerical simulations. 

The force expressions in Supplementary Equations~(\ref{eq:Fg_SI}), (\ref{eq:Fc_SI}), (\ref{eq:stokes_SI}), and~(\ref{hymo_SI}) allow us to simulate the motion of each beating particle $i$ with Newton's second law:
\begin{equation}
    \frac{d\mathbf{v}_i}{dt} = \frac{1}{m_{\text{eff}}}(\mathbf{F}_{\text{g},i}+\mathbf{F}_{\text{c},i}+\mathbf{F}_{\text{d},i})
\end{equation}
We followed \cite{Hauser2018} in introducing a scaling factor for the effective mass ($m_{\text{eff}}=\Lambda_{\text{m}}m$) to account for the added mass of liquid experienced during particle acceleration. The two fitted parameters of the model, $\Lambda_{\text{m}} = 11.25$ and $\Lambda_{\text{d}} = 0.35$, were kept constant across simulations of different H$_2$O$_2$ concentrations. The model outlined above was solved numerically again with MATLAB’s ode45 Runge-Kutta solver. As all three forces are also dependent on the instantaneous bubble volume (equivalently, the radius), we zoom in to the catalytic surface and study the reaction kinetics as the final piece of the puzzle.

The volume of the O$_2$ bubble as a function of time, $V_{\text{b}(t)}$, is dictated by the rate of O$_2$ generation, which in turn is dependent on the free platinum patch surface area $A_{\text{Pt,free}}$, as well as the peroxide concentration [H$_2$O$_2$]. For a given experiment, we assume that the peroxide is in excess and [H$_2$O$_2$] is a constant throughout, based on the absence of a shift in the beating frequency (Supplementary Fig.~\ref{fig:windowed_recurrence}). A well-studied catalytic reaction, the decomposition kinetics of H$_2$O$_2$ on noble metal and oxide surfaces can be described by the classic Langmuir-Hinshelwood mechanism~\cite{Plauck2016,Lin1998}:
\begin{equation}
    \frac{d{{V}_{\text{b}}}}{dt}=\frac{k{{A}_{\text{Pt,free}}}({{V}_{\text{b}}})[{{\text{H}}_{\text{2}}}{{\text{O}}_{\text{2}}}]}{1+{{K}_{\text{H}}}[{{\text{H}}_{\text{2}}}{{\text{O}}_{\text{2}}}]}
    \label{eq:langmuir_SI}
\end{equation}
where $k$ is a constant encompassing the reaction rate constant, the specific volume of O$_2$, and the areal density of the surface sites. The kinetic equation represents that the rate is first order with respect to the concentration of bound surface sites, which are saturated at increasing peroxide concentration modulated by the binding constant $K_{\text{H}}$. In the single particle scenario, $A_{\text{Pt,free}}$ decreases over time as the bubble underneath the particle starts to limit the accessible catalytic surface area. This leads to a reduced $dV_{\text{b}}/dt$ and therefore a self-limiting reaction. Inspection of the 2-particle beating videos, on the other hand, shows a near-linear increase of the bubble volume up until the moment of merger. This observation suggests that the bubbles in the beating system do not grow beyond the critical $V_{\text{b}}$ which marks the onset of catalytic surface blockage, and that $A_{\text{Pt,free}}\approx A_{\text{Pt}}$. The resultant time-independent reaction rates at different H$_2$O$_2$ molarities were fitted to the Langmuir-Hinshelwood kinetics, outputting $k=3.025\times10^{-10}$m$^4$s$^{-1}$mol${^{-1}}$ and $K_{\text{H}}=0.677$L/mol. This parameterised kinetics was used in our quantitative model to account for the effects of H$_2$O$_2$ concentration and Pt surface area (e.g. in Fig. 1j of the main text). 

All parameters used in the mechanistic model are listed in Supplementary Table~\ref{table:modPar_SI}.

% START OF TABLE

\begin{table}[h!]
\centering
\caption{\textbf{Parameters used in the mechanistic model.}}
\label{table:modPar_SI}
\begin{tabular}{clc}
\multicolumn{1}{l}{}                          &                                                                                         & \multicolumn{1}{l}{}            \\ 
\hline
Symbol                                        & \multicolumn{1}{c}{Parameter}                                                           & Conventional Unit               \\ 
\hline
$\beta$                                       & Curvature at the apex of the drop                                                       & {[}m$^{-1}$]                      \\
$\gamma$                                      & Interfacial tension of the liquid-air interface                                         & {[}N/m]                         \\
$\gamma_\text{c}$                             & Capillary constant of the liquid drop                                                   & {[}m$^{-2}$]                      \\
$\theta$                                      & Angle of the liquid-air interface relative to the lateral dimension                     & {[}rad]                         \\
$\theta_\text{c}$                             & Three phase contact angle                                                               & {[}rad]                         \\
$\lambda$                                     & Dimensionless distance between two floating objects used in hydrodynamic mobility       & {[}-]                           \\
$\Lambda_\text{b}$                            & Volume fraction of gas bubble lying below the undisturbed interface                     & {[}-]                           \\
$\Lambda_\text{d}$                            & Effective drag scaling factor                                                           & {[}-]                           \\
$\Lambda_\text{m}$                            & Effective mass scaling factor                                                           & {[}-]                           \\
$\mu$                                         & Dynamic viscosity of the liquid                                                         & {[}N$\cdot$s$\cdot$m$^{-2}$]        \\
$\rho$                                        & Lateral coordinate relative to the drop's apex                                          & {[}m]                           \\
$\rho_\text{a}$                               & Density of air                                                                          & {[}kg/m$^3$]                    \\
$\rho_\text{l}$                               & Density of liquid                                                                       & {[}kg/m$^3$]                    \\
$\Sigma$                                      & Buoyancy-corrected dimensionless weight                                                 & {[}-]                           \\
$A_\text{Pt}$                                 & Surface area of the Pt patch                                                            & {[}m$^2$]                         \\
$A_\text{Pt,free}$                            & Accessible area of the Pt patch                                                         & {[}m$^2$]                         \\
$B$                                           & Bond number                                                                             & {[}-]                           \\
$F_\text{c}$                                  & Lateral component of the local capillary force between floating objects                 & {[}N]                           \\
$F_\text{d}$                                  & Lateral component of the drag force                                                     & {[}N]                           \\
$F_\text{g}$                                  & Lateral component of the global buoyancy-corrected gravitational force towards the apex & {[}N]                           \\
$g$                                           & Gravitational acceleration~                                                             & {[}m/s$^2$]                       \\
$G$                                           & Hydrodynamic mobility                                                                   & {[}-]                           \\
{[}H$_2$O$_2$]                                & Concentration of hydrogen peroxide                                                      & {[}mol/L]                       \\
$k$                                           & Rate constant of peroxide decomposition per unit area~                                  & {[}m$^4\cdot$s$^{-1}\cdot$mol$^{-1}${]}  \\
$K_\text{H}$                                  & Surface binding constant~of peroxide decomposition                                      & {[}L/mol]                       \\
$K_n$                                         & Modified Bessel function of the second kind of order $n$                                & {[}-]                           \\
$l$                                           & Lateral coordinate from the centre of a floating object                                 & {[}m]                           \\
$L_\text{c}$                                  & Capillary length                                                                        & {[}m]                           \\
$m$                                           & Mass of a microparticle                                                                 & {[}kg]                          \\
$m_\text{eff}$                                & Effective mass of a microparticle                                                       & {[}kg]                          \\
$R$                                           & Radius of a floating object                                                             & {[}m]                           \\
$R_\text{b}$                                  & Instantaneous radius of a bubble                                                        & {[}m]                           \\
$s$                                           & Coordinate along the liquid-air interface relative to the drop's apex                   & {[}m]                           \\
\begin{tabular}[c]{@{}c@{}}$t$\\\end{tabular} & Time                                                                                    & {[}s]                           \\
\begin{tabular}[c]{@{}c@{}}$v$\\\end{tabular} & Lateral component of a microparticle's velocity                                         & {[}m/s]                         \\
\begin{tabular}[c]{@{}c@{}}$V$\\\end{tabular} & Integral volume used in calculating the drop profile                                    & {[}m$^3$]                         \\
$V_\text{b}$,~$V'_\text{b}$                   & Volume of a bubble                                                                      & {[}m$^3$]                         \\
$V_\text{drop}$                               & Volume of the drop                                                                      & {[}m$^3$]                         \\
$z$                                           & Vertical coordinate relative to the drop's apex                                         & {[}m]                          
\end{tabular}
\end{table}
% END OF TABLE

\pagebreak

\subsection{Thermodynamics of Particle Beating}
\label{sec:model_SI}
Asymmetry-induced order is a process by which explicit symmetry-breaking (spontaneous or otherwise) leads to the emergence of ordered states in a system~\cite{Medeiros2021,Nicolaou2021,Zhang2021_sync,Zhang2017_asymmetry}. Hence, asymmetry-induced order requires both a symmetry whose breaking can be observed, and a clear notion of ``degree of order.'' Which symmetry to break is inherently a system-dependent question, and as such there are no general means of choosing between symmetry groups to achieve a desired outcome. However, the so-called degree of order of a system is a challenging property to formally specify in general. For one, what is meant by order is often ill-defined or underspecified. Secondly, even when provided with a means to metricize order, such metrics are often analytically and computationally intractable because they require global knowledge of system states---as is the case for calculating entropy. This is further complicated by the fact that, far from equilibrium, entropy is not sufficient to establish the robustness, stability, or persistence of system configurations (all of which are attributes often ascribed to ``orderly'' states)~\cite{Landauer1975}. To this end, physicists have made use of order parameters to establish more narrowly-construed notions of order on a case-by-case basis for particular systems~\cite{Vicsek1995,Parisi1983}. 

Recent work in nonequilibrium thermodynamics has made strides towards describing the emergence of order more generally in broader classes of complex systems. Rattling theory is a novel thermodynamic theory describing the emergence of order and self-organization in ``messy'' nonequilibrium dynamical systems~\cite{Chvykov2021, Chvykov2018}. The success of rattling theory depends crucially on the definition of the class of systems it considers to be messy. The rattling ansatz sees the behavior of complex systems as stochastic diffusion processes taking place in high-dimensional configuration spaces in the presence of energy influxes driving them out of equilibrium. Any system whose behavior can be described by such configuration-space diffusion falls under the class of messy systems described by rattling theory. Modelling the behavior of systems as diffusion processes is what enables an analytical determination of nonequilibrium steady-state density, and, as a consequence, an understanding of self-organization. Empirically, this approach has been shown to predict the long-term behavior of a wide variety of systems, from canonical chaotic systems~\cite{Jackson2021} to swarms of robots~\cite{Chvykov2021}, and is expected to apply across diverse active matter systems as well~\cite{Palacci2013,Corte2008}. 

At the heart of the theory lies a precise, local, and computable measure of order, \textit{rattling}, from which the theory derives its name. Rattling measures the way in which system configurations respond to external force fluctuations: Rapid, uncorrelated configurational changes produce high rattling values, and slow, correlated changes produce low rattling values. When a system's response to local force fluctuations is random (i.e., has Gaussian statistics), rattling is exactly the entropy of its configurational velocities. As a quantity, the rattling $\mathcal{R}(q)$ of a system at configuration $q$ is
\begin{equation}
    \mathcal{R}(q) = \frac{1}{2}\log \det \langle \dot{q}_i(t), \dot{q}_j(t) \rangle_{q(0)=q}
    \label{eq:rattling_SI}
\end{equation}
where $\langle \cdot, \cdot \rangle$ is the covariance tensor of the system's configurational velocities (i.e., two-point correlation function) averaged over an ensemble of dynamical trajectories initialized at $q$. Using this definition, we can state the central prediction of rattling theory, known as the ``low-rattling selection principle.'' The principle expresses a relationship between the magnitude of system-level fluctuations measured at a particular configuration (i.e., rattling $\mathcal{R}(q)$), and said configuration's prevalence in the system's nonequilibrium steady-state density.  In particular, this relationship is of Boltzmann-like form:
\begin{equation}
    p(q) \propto e^{-\gamma \mathcal{R}(q)}
    \label{eq:lowrattselect_SI}
\end{equation}
where $\gamma$ is a constant of order 1. This relationship shows that configurations with remarkably low entropy dynamical responses (i.e., low rattling) are exponentially preferred in the system's steady-state. Thus, rattling and its associated selection principle are able to sufficiently establish the robustness, stability, and persistence of the configurations of complex systems.

In summary, rattling captures the ways in which correlations among disorderly degrees of freedom give rise to system-level fluctuations of different magnitudes. Then, the low-rattling selection principle states that such system-level fluctuations bias the nonequilibrium steady-state of a complex system towards configurations in which the system experiences remarkably low magnitude fluctuations. Furthermore, this spontaneous selection of low rattling configurations necessarily requires that strong correlations between degrees of freedom arise, and thus for orderly behaviors to emerge. Interested readers looking for a complete treatment of this material, as well as theoretical derivations and experimental validation, are referred to~\cite{Chvykov2021}. Equipped with a precise way to quantify order in a broad class of complex systems, we may now develop a system-specific understanding of the ways in which symmetry-breaking affects the rattling of our system of beating particles in hopes of finding strategies to stabilize periodic system beating for $N>2$.

In order to elucidate the role that symmetry-breaking may play in the self-organized states of our system of active microparticles, we must now consider specific system symmetries and their relationship to the magnitude of system-level fluctuations. While our system is not invariant to the action of any obvious continuous symmetry groups, it is permutation-symmetric~\cite{Ferre2015}. This is to say that our collection of microparticles are all dynamically identical (up to fabrication tolerances). Hence, one promising avenue to investigate is the different ways in which permutation-symmetry breaking may lead to order in our system. Based on results from our mechanistic modelling of particle beating, we know that there are two ways in which the dynamics of individual microparticles can be made distinct from one another. First, we know that changing the volumetric shape of particles will lead to different local hydrodrynamic drag properties. Second, we know that changing the buoyancy of particles also produces local changes to individual microparticle dynamics through its effect on capillary forces.  However, changing the shape of our microparticles requires major changes to their fabrication, as well as nontrivial modifications to the mechanistic model. In contrast, we can easily modify a particle's buoyancy by modulating the volume of the bubble forming underneath the particle, which we can in turn control through the size of their Pt patch.

To explore the role of permutation-symmetry breaking on our system, we constructed a simple model that we can work with analytically from the perspective of rattling theory. In line with the rattling ansatz, our model considers the configurational dynamics of collectives of beating particles as a diffusion process. We incorporate the effect of heterogenous particle buoyancies through the inclusion of a parameter modulating the size of bubbles in analogy to the role of the Pt patch. Our beating particles are perfectly suited for this sort of analysis, even more so than others (e.g., robot swarms). In part, this is due to the physics of fluid dynamics at low-Reynolds numbers ($\sim$0.25 Re for our system)~\cite{Purcell1977}. In this regime, inertia ceases to influence the behavior of systems, leaving viscous forces and stochastic thermal fluctuations to affect their dynamics substantially---thereby making a diffusive approximation natural.

Our model elucidates the role of design parameters on the structure of the system-level fluctuations on the basis of two primary assumptions. First, we assume that the behavior of each individual particle $i$ is monotonically modulated by some real-valued design parameter $U_i$ from a set $U=\{U_1, \cdots, U_N\}$ for a system of $N$ particles. These design parameters correspond by analogy to the Pt patch size. Second, we assumed that particle $i$'s bubble burst only affects the other members of the collective and not itself, which broadly matches experimental observations. We can think of the $U_i$ parameters as implicitly determining the strength of the impulse imparted by particle $i$'s bubble burst onto its neighbors. In particular, we model the effect of this parameter and the bubble burst strength $a_i$ according to the following Boltzmann-like monotonic relationship, 
\begin{equation}
    a_i = \frac{1}{Z} e^{-U_i}
    \label{eq:probburst_SI}
\end{equation}
where $Z$ is a normalization factor given by $Z = \sum_{i=1}^{N}e^{-U_i}$. In other words, the $a_i$ parameters can be thought of in analogy to the size (and strength) of bubbles that a given particle can support. Hence, we can motivate this modeling choice by envisioning the gas in bubbles distributing itself according to an energy landscape specified by our $U_i$ parameters, and thusly influencing the bubble popping strength $a_i$. The normalization factor $Z$ arises from the fact that we are not interested in the absolute magnitude of the bubble bursts but rather the effect of their relative magnitudes on the collective behavior. 

In the main text, we made use of an observable termed the ``breathing radius'' for purposes of analysis. The breathing radius is the mean Euclidean distance of the particles to the centroid of the collective. Similarly, here we will only consider the statistical properties of the dynamics of a breathing-radius-like observable, $\bar{r}(t)$, under a simple diffusive model. As in the main text, $\bar{r}(t)$ is an averaged quantity over particles: $\bar{r}(t) = \sum_{i=1}^{N} r_i(t)$. By assumption, a bubble burst at particle $i$ leaves particle $i$ stationary, but a burst from some neighbor $j$ exerts an impulse of random direction onto particle $i$. In this case, the dynamics of $r_i(t)$ evolve according to
\begin{equation}
    \dot{r}_i(t) = \sum_{j\neq i} a_j \cdot \xi_j
    \label{eq:modeldyn_SI}
\end{equation}
where $\xi_j$ is normally-distributed delta-correlated multiplicative noise in the It\^{o} convention. Note that this construction results in an anisotropic diffusion tensor without spatial dependence, as we are not modelling the geometry of interparticle interactions but rather their statistical fluctuations. From this specification of the system's diffusive dynamics, we can apply rattling theory to understand the effect of our design parameters $U_i$ on the self-organized collective behavior of the system.

\begin{figure*}[t]
    \centering
    \includegraphics[width=0.6\linewidth]{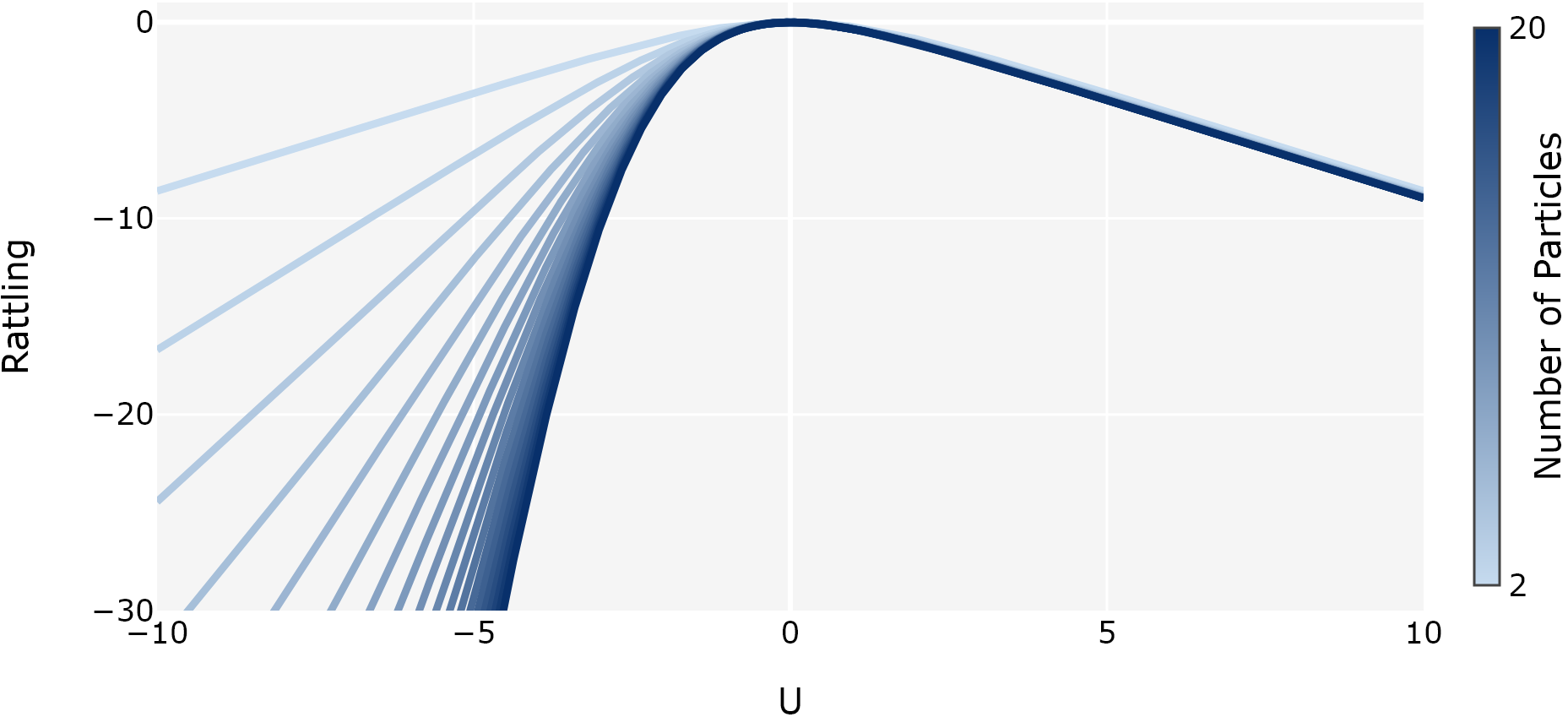}
    \caption{\textbf{Rattling as a function of patch size in diffusive model.} Here, we study the effect of a given particle's $U$ parameter (in analogy to Pt patch size) on the rattling of collectives of varying sizes. Note that we subtract the constant offset in rattling due to system size so that $\mathcal{R}=0$ at $U=0$ for all $N$. We find that any variability in the size of the particle's patch produces a drop in rattling, leading to asymmetry-induced order. When a particle becomes inert as $U$ increases, it stops contributing to system-level fluctuations, leading to a modest drop in rattling independent of $N$. However, as $U$ decreases the modified particle's bubble bursts dominate and effectively become the sole source of variance in the system's configurational degrees of freedom. Such coordination among degrees of freedom leads to a sharp drop in rattling dependent on $N$.}
    \label{fig:DLmodel_SI}
\end{figure*}

Given this formulation of the system dynamics, we proceed by calculating the effect of parameter changes on the magnitude of system-level fluctuations. Letting $r(t)=[r_1(t), \cdots, r_N(t)]^T$, the correlation structure of the system is
\begin{equation}
    \langle \dot{r}_i(t), \dot{r}_j(t) \rangle =  \sum_{k \neq i} \sum_{l \neq j} a_k a_l \delta_{kl} = \sum_{k\neq i, j} a_k^2 = \frac{1}{Z^2} \sum_{k\neq i, j} e^{-2U_k}
    \label{eq:covariance_SI}
\end{equation}
where $\delta_{kl}$ is the Kronecker delta. We note that the correlation structure of the system has no dependence on time (i.e., it has infinite temporal correlation) and no dependence on configuration $r(t)$, leaving the design parameters $U_i$ as the only variables with an effect on the system behavior. Finally, in order to express the system's rattling in terms of its design parameters we require an analytical expression for the determinant of its covariance tensor, which is challenging in general. Fortunately, for this particular correlation structure there exists such a closed-form expression, which enables us determine the system's rattling as a function of its parameters:
\begin{equation}
    \mathcal{R}(U) = \frac{1}{2}\log \det \langle \dot{r}_i(t), \dot{r}_j(t) \rangle = \log \left( \frac{(N-1)\prod_{i=1}^{N}e^{-U_i}}{Z^N}\right).
    \label{eq:ratt_analytical_SI}
\end{equation}

Equipped with an understanding of how the system's parameters affect its rattling (and thus its degree of order), we can now use the model as a tool to guide our experimental design. While there are infinitely many parameter combinations for a given collection of $N$ particles, one of the simplest design alterations to study is the effect of a single particle differing from the rest---for reasons that we will see shortly, we term this particle a \textit{designated leader}. In this setting, one particle will have its parameter be $U_{DL}$ while the rest of the $N-1$ particles will have it be $\bar{U}$ (which we take to be a constant fixed a priori). Rearranging the expression in Supplementary Equation (18), we have the following expression
\begin{equation*}
    \mathcal{R}(U_{DL},N) = -U_{DL}+\log \left( \frac{(N-1)e^{-(N-1)\bar{U}}}{\left(e^{-U_{DL}}+(N-1)e^{-\bar{U}}\right)^N}\right),
    \label{eq:ratt_analytical_SI_DL}
\end{equation*}
which allows us to make predictions about the behavior of a collection of $N$ beating particles with a single designated leader. 

However, much in the same way that entropy can trivially depend on system size (e.g., number of microstates), our expression for rattling in Supplementary Equation~(\ref{eq:ratt_analytical_SI_DL}) does as well. Thus, to focus on the dependence of $\mathcal{R}(U_{DL},N)$ on $U_{DL}$, we subtract the constant bias that system size contributes to the value of rattling. To do this, we calculate $\mathcal{R}(U_{DL},N)-\mathcal{R}(\bar{U},N)$ for a choice of $\bar{U}$ that we fix across all system sizes, where we note that $\mathcal{R}(\bar{U},N)$ is merely a constant that offsets the value of rattling to be zero when $U_{DL}=\bar{U}$. Since $\mathcal{R}(\bar{U},N)$ is exclusively a function of the number of particles for a given $\bar{U}$, subtracting it from $\mathcal{R}(U_{DL},N)$ precisely removes the constant contribution of system size to the overall magnitude of rattling. As detailed in~\cite{Chvykov2021}, constant offsets to the rattling values of a system do not affect its behavior. Only changes to the rattling landscape---that is, changes to the relative rattling values between configurations (or parameters)---have an effect on system behavior. This implies that comparing the absolute rattling values across systems is of limited use, which motivates our approach (as in Fig.~2d and Supplementary Fig.~\ref{fig:DLmodel_SI}).

In Supplementary Fig.~\ref{fig:DLmodel_SI} we show the results of varying the parameters of the designated leader for collectives of various sizes, while fixing $\bar{U}=0$ and subtracting the bias in rattling due to system size. Crucially, we observe that any deviation from the parameter values of the rest of members of the collective (i.e., away from $U_{DL}=0$) results in a reduction in rattling. Thus, our model predicts that any amount of heterogeneity will lead to increasingly ordered system states. Such asymmetry-induced order has been studied in networked systems of oscillators~\cite{Zhang2017_asymmetry,Zhang2021_sync,Nicolaou2021,Medeiros2021}, but its emergence as a low-rattling phenomenon is a novel finding.

Through this mechanism, order arises in one of two distinct ways. First, as $U_{DL}$ increases, the designated leader particle becomes effectively inert. This is to say that the strength of its bubble bursts $a_{DL}$ asymptotically approach zero, as though it were a patchless particle. As a result, the leader particle acts as dead weight and does not contribute to system-level fluctuations, leading to a modest decrease in rattling---independent of the total number of particles---that matches experimental observations. Second, as $U_{DL}$ decreases, the designated leader particle's bubble bursts become stronger and its contribution to the magnitude of system-level fluctuations dominates over those of other particles. In turn, this effectively leads to a concentration of all variability and randomness in the system into a single of its many degrees of freedom, thereby leading to significant correlations in the behavior of all particles and a resulting drop in rattling. Note that as more particles are added more degrees of freedom become correlated, leading to sharper drops in rattling as a function of $N$. Hence, on the basis of these results and other studies of asymmetry-induced order we chose to study the influence of designated leaders on the collective behavior experimentally by producing leader particles with larger Pt patches. 

\begin{figure*}[t]
    \centering
    \includegraphics[width=0.85\linewidth]{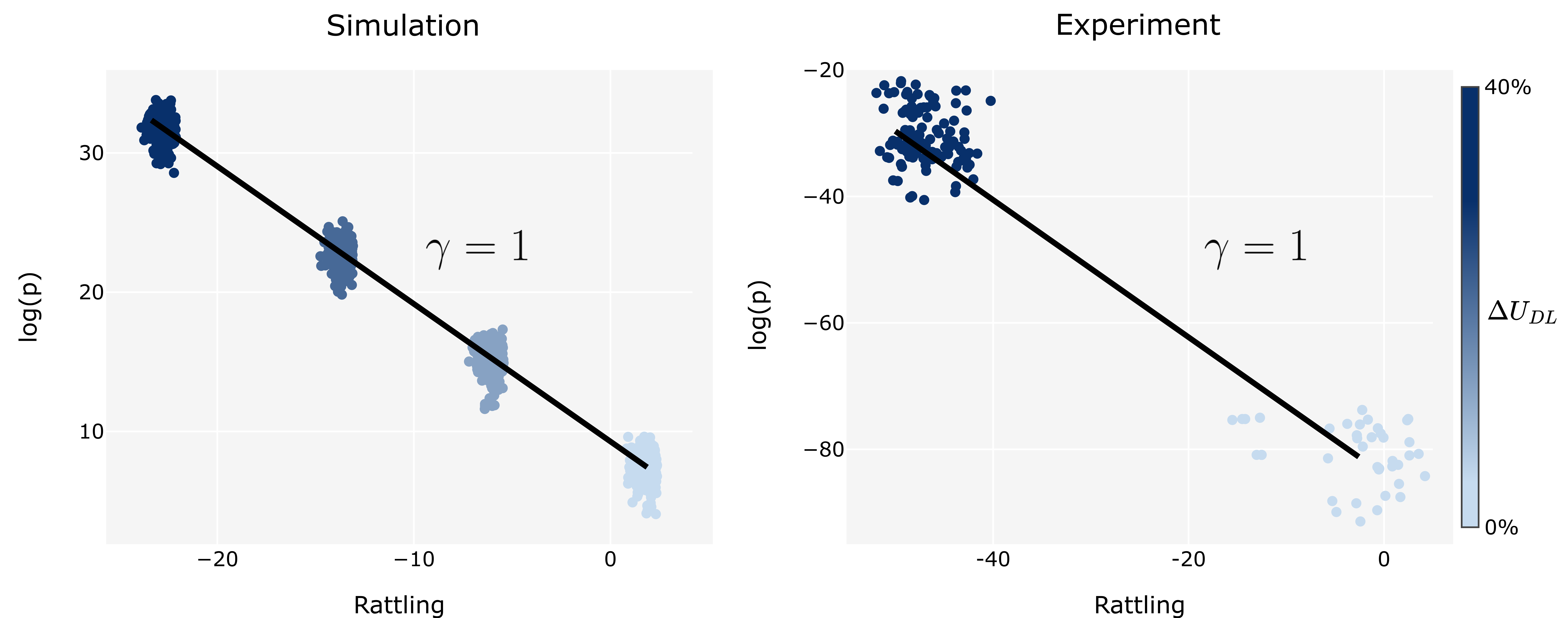}
    \caption{\textbf{Effect of designated leader on self-organization.} By introducing a designated leader, the entropy of the bubble burst forcing patterns decreases (since they become periodic), which has an effect on the self-organization of the system. On the left panel, we simulate the dynamics in Supplementary Equation~(\ref{eq:modeldyn_SI}) and calculate their rattling and steady-state densities numerically. On the right panel, we consider experimental data from an 8 particle collective in both standard ($\Delta U_{DL}=0\%$) and designated leader configurations ($\Delta U_{DL}=40\%$), which we then process using the same procedure as for the left panel. While the absolute magnitudes of parameter values for the simulation are arbitrary, the $\Delta U_{DL}$ values are determined from the actual Pt patch sizes used on the experimental systems. For both the simulated and the experimental data, the results are consistent with rattling theory (in particular, Supplementary Equation~(\ref{eq:lowrattselect_entropy_SI}) with $\gamma=1$). This procedure was then repeated for experimental collectives of other sizes with the same results. Hence, bubble burst patterns of varying entropy (which depend on system design parameters) provide an explanation for the emergence of system order that is consistent with our results.}
    \label{fig:DLcomparison_SI}
\end{figure*}

Another consequence of applying the rattling ansatz to the behavior of our collective of beating particles is that it allows one to reinterpret the relationship between system elements. In the theory, configurations with exceptionally orderly responses to external driving forces are selected for the nonequilibrium steady-state of complex systems. Similarly, we can think of the relationship between the particle configurations (i.e., their relative positions and orientations) and the sequence of forces the system experiences due to bubble bursts according to this dichotomy. Working from this perspective, rattling theory then suggests that the entropy of the sequence of bubble bursts can prevent the system from finding orderly configurations by increasing their rattling (see~\cite{Chvykov2021}, and in particular figure~4 within). More precisely, we have
\begin{equation}
    p(q) \propto e^{-\gamma(\mathcal{R}(q)+ S(q))},
    \label{eq:lowrattselect_entropy_SI}
\end{equation}
where $S(q)$ is the entropy of the driving forces affecting the system at configuration $q$. Importantly, this expression shows that the effect of drive entropy is to simply offset a configuration's rattling.

In the main text, we observed that our system design parameters (i.e., the particle Pt patch sizes) do have a profound effect on system behavior and also on the bubble burst sequence---changing its behavior from seemingly random to almost perfectly periodic (see Fig.~2). If we were to accept the hypothesis presented by Supplementary Equation~(\ref{eq:lowrattselect_entropy_SI}), then this difference in behavior should be explained by the difference in the entropy generated by the bubble burst patterns at different system parameters. Moreover, if this is the case, then the results from analyzing standard and designated leader systems should lie on the log-linear correlation of Supplementary Equation~(\ref{eq:lowrattselect_entropy_SI}), with a slope of $\gamma$ (which nominally is of order 1). Indeed, this is precisely what we observe in Supplementary Fig.~\ref{fig:DLcomparison_SI} for simulations of the model dynamics and for our experimental data samples. 

While throughout this section we have motivated the analytical model in specific reference to our experimental beating particles, our model and results generalize beyond our system. At its core, our generic model describes the structure of statistical fluctuations in a collection of strongly interacting degrees of freedom (i.e., under strong mixing conditions). This is to say that the details of how the magnitudes of said fluctuations are parametrized by system properties are inessential to the results. Particularly, the equation for rattling in Supplementary Equation~(\ref{eq:ratt_analytical_SI}) can be expressed in terms of $a_i$ directly as $\log\big((N-1)\prod_{i=1}^{N}a_i\big)$, which allows one to freely model the way in which individual degrees of freedom contribute to the overall system-level fluctuations. Thus, rattling theory, as well as the mechanism we have outlined for asymmetry-induced order, present a general framework from which to understand the effect of system design parameters on the self-organized behaviors of the system---providing a novel approach to micro-system design based on thermodynamic principles relevant to the scales of interest.

\pagebreak

\subsection{Note on Microelectronic Low-Frequency Oscillators}
\label{sec:lfo_SI}
In this section, we elaborate on the design and fabrication challenges of microelectronic oscillators with a frequency on the order of a hertz, which we briefly alluded to in the Introduction part of the main text. Given the relatively large footprints of integrated capacitors and inductors available, RC- and LC-based oscillators are hardly compatible with the limited space on micrometre-sized machines~\cite{Funke2016}. For example, the frequencies of RC oscillators, such as a bi-inverter or a Schmidt Trigger oscillator, are on the order of the reciprocal of their respective RC constants, i.e. $f_{\text{RC}}\sim\mathcal{O}(1/RC)$. Taking the capacitance to be a generous 40pF for an area of 100$\mu$m$\times$100$\mu$m~\cite{Molnar2021}, one would require a massive resistor of 25G$\Omega$ to achieve an RC time constant of 1s. Assuming a resistivity of 100k$\Omega$/$\mu$m$^2$ of polysilicon, this resistor alone would occupy 2.5$\times10^5\mu$m$^2$. Alternatively, one may opt to use a frequency divider to bring the kHz-order frequency of a typical microelectronic relaxation oscillator down to 1Hz. Suppose the starting frequency is 17kHz~\cite{Lee2018_freqdivider}, a cascade of 15 T flip-flops is needed, each of which is constructed from at least 20 transistors~\cite{Harris2021}. Should 300 transistors be fabricated onto a 100$\mu$m$\times$100$\mu$m microchip, the appropriate transistor node would be 500nm. While well within the realm of possibility, such technology typically still requires the involvement of a commercial foundry outside of academic institutions. Similarly, thyristor-based oscillators of frequencies from 20Hz and up have been foundry-fabricated with a feature size of 180nm~\cite{Funke2016}. The integrated circuit design expertise and capital investment required are the reasons for a high barrier-to-entry. Note that the area reserved for onboard energy harvesting and storage units, as well as for miscellaneous electronics, may further constrain the real estate available to the microelectronic oscillator.

\subsection{Note on the Fuel Cell's Open-Circuit Voltage}
\label{sec:ocv_SI}
Supplementary Fig. 13 shows that the open-circuit voltages of the Pt-Au and Pt-Ru fuel cell devices, $V_{\text{OC}}$, exhibits a very weak dependence on the peroxide concentration [H$_2$O$_2$], unlike the trend of the short-circuit current densities (Fig. 4c and Supplementary Fig. 14). Here we provide a simple explanation based on electrochemical kinetics. We consider the following two pairs of forward and reverse reactions taking place on a single electrode: 

\begin{chemmath}
    \left. \begin{aligned}
     & H_2O_2 \reactrarrow{0pt}{5mm}{}{} O_2+2H^{+}+2e^{-} \\ 
     & O_2+2H^{+}+2e^{-} \reactrarrow{0pt}{5mm}{}{} H_2O_2
    \end{aligned} \right\}{{R}_{1}, \text{ }\phi_{\text{eq,1}}^{\circ}}=+0.68\text{V}
    \label{eq:r1_SI}
\end{chemmath}
\begin{chemmath}
    \left. \begin{aligned}
      & 2H_2O \reactrarrow{0pt}{5mm}{}{} H_2O_2+2H^{+}+2e^{-} \\ 
     & H_2O_2+2H^{+}+2e^{-} \reactrarrow{0pt}{5mm}{}{} 2H_2O
    \end{aligned} \right\}{{R}_{2}, \text{ }\phi_{\text{eq,2}}^{\circ}}=+1.77\text{V}
    \label{eq:r1_SI}
\end{chemmath}where $\phi_{\text{eq}}^{\circ}$ denotes the standard equilibrium potentials. The Butler–Volmer equation suggests that only one half-reaction from R$_1$ and R$_2$ each is dominant at the mixed potential $\phi_{\text{mix}}$, defined as the potential where the total current equals 0~\cite{Park2013}. If we consider the oxidative half-reaction of R$_1$ and the reductive half-reaction of R$_2$ (choosing the other two half-reactions does not alter the conclusion), the full Butler-Volmer kinetic expression is given by~\cite{Corbin2019}:
\begin{align}
    i_1(\phi)&=nFk_1[\text{H}_2\text{O}_2]^{\nu_{\text{H}_2\text{O}_2}}\exp{\Big[\frac{\alpha_1F}{\bar{R}T}\phi\Big]}\label{eq:bv1_SI}\\
    i_2(\phi)&=-nFk_2[\text{H}_2\text{O}_2]^{\nu_{\text{H}_2\text{O}_2}}[\text{H}^{+}]^{\nu_{\text{H}}}\exp{\Big[-\frac{(1-\alpha_2)F}{\bar{R}T}\phi\Big]}\label{eq:bv2_SI}
\end{align}
where $\phi$ is the applied potential on the absolute scale, $i_1(\phi)$ and $i_2(\phi)$ the respective current densities, $n$ the number of electrons transferred, $F$ the Faraday constant, $k$ the rate constants, $\nu$ the reaction orders, $\alpha$ the the transfer coefficients, $\bar{R}$ the universal gas constant, $T$ the absolute temperature. We can obtain the mixed potential $\phi_{\text{mix}}$ by solving:
\begin{equation}
    i_1(\phi_{\text{mix}})+i_2(\phi_{\text{mix}})=0
\end{equation}
which is equivalent to $-i_1(\phi_{\text{mix}})/i_2(\phi_{\text{mix}})=1$. While the exact form of the solution is of little relevance to us, the division of the right-hand side of Supplementary Equation~(\ref{eq:bv1_SI}) by that of Supplementary Equation~(\ref{eq:bv2_SI}) reveals the cancellation of the [H$_2$O$_2$] terms under the typical assumption of equal reaction order. That is, $\phi_{\text{mix}}$ is independent of the peroxide concentration for a given electrode. Because the open-circuit voltage between two spatially separated electrodes (such as Pt and Ru) is essentially the difference in the respective mixed potentials ($\Delta\phi_{\text{mix}}$), $V_{\text{OC}}$ naturally sees little dependence on [H$_2$O$_2$]. This allows us to compare our $V_{\text{OC}}$ measurements with past mixed potential studies carried out at lower  [H$_2$O$_2$]. For example, Wang and colleagues~\cite{Wang2006} measured a $\Delta\phi_{\text{mix}}$ of 30mV between Pt and Au, and 140mV between Pt and Ru, both consistent with our results.

\subsection{Note on the Energy Expenditure}
\label{sec:effiency}
\subsubsection{Energy Conversions of the Mechanical Oscillation}

Within each period of the emergent mechanical oscillation, chemical energy stored in the H$_2$O$_2$ fuel is converted into the particles' kinetic energy upon the collapse of the O$_2$ bubble. The kinetic energy imparted to two outgoing particles simply take the form of $E_{\text{k}}= m v^2$, where $m$ is the mass of each particle and $v$ the maximal velocity right following the bubble collapse. With $m=2.34\mu$g for a 500$\mu$m-diameter particle and $v=3.2\times10^4\mu$m/s measured from experiments, $E_{\text{out}}$ is estimated to be $2.40\times10^{-12}$J per cycle. \\

\noindent The chemical energy consumed per cycle may be computed as:
\begin{equation*}
    E_{\text{chem}}=\frac{2PV_{\text{b,th}}\Delta H}{RT}
\end{equation*}
where $V_{\text{b,th}}$ denotes the bubble volume at threshold, estimated to be 9.81$\times 10^{-2}\mu$L in a 2-particle homogeneous system in 1mL of 10\% H$_2$O$_2$. We assume an ambient pressure $P$ of 1atm and temperature $T$ of 25$^{\circ}$C, as the excess Laplace pressure within the bubble before collapse is a negligible $5.0\times10^{-3}$atm. $\Delta H$, the enthalpy change of the decomposition reaction, is 98.24kJ/mol at given conditions, equivalent to an energy density of 2.89kJ/g H$_2$O$_2$ or 0.29kJ/g 10wt\% H$_2$O$_2$ solution \cite{Wehner2016}. $E_{\text{chem}}$ per cycle is computed to be $7.88\times10^{-4}$J. The portion of the chemical energy converted to the work of expansion is:
\begin{equation*}
    W_{PV}=P_{\text{atm}}V_{\text{b,th}}+4\pi\gamma R_{\text{b,th}}^2
\end{equation*}
where $R_{\text{b,th}}$ is the threshold radius assuming a spherical bubble. The latter term of $7.41\times10^{-8}$J is the surface energy $E_{\text{surf}}$, i.e. the work against the Laplace pressure during bubble growth. To summarize, therefore, 1.26\% of the original chemical energy contributes to a $W_{PV}$ of $1.00\times10^{-5}$J. 0.74\% of the work of expansion is stored as the surface energy. Finally, the kinetic energy gained by the particles account for 0.032{\textperthousand} of surface energy stored in the bubble. 

\subsubsection{Energy Conversions of the Microgenerators}
As the microgenerator converts the chemical energy from H$_2$O$_2$ decomposition to electrical work, it is of interest to calculate the proportion of total H$_2$O$_2$ molecules consumed which contributed to the electrical current \cite{Wang2013_efficiency,Paxton2006}. Given that each electrochemically redoxed H$_2$O$_2$ molecule transfers an electron, ON-state currents of 180.66nA (in the absence of an electrical load) and 15.27nA (with a load, i.e. the actuator) are respectively attributed to $1.87\times 10^{-12}$ and $1.58\times10^{-13}$ moles of H$_2$O$_2$ per second. These correspond to 0.76{\textperthousand} and 0.063{\textperthousand} of the total peroxide consumption rate ($2P/RT\cdot dV_{\text{b}}/dt=2.45 \times 10^{-9}$mol/s), respectively. The former is in agreement with prior literature \cite{Wang2013_efficiency}, which estimated an electrochemical contribution of 0.5{\textperthousand}. Since more than 99.9\% of the consumed H$_2$O$_2$ decompose via the same non-electrochemical pathway as in the beating particles with no fuel cells aboard, generation of the electrical current has a negligible impact on the mechanical oscillation if all other conditions are kept the same. Along the same lines, additional fuel cell particles are not expected to diminish the electrical signals observed.

\clearpage
\newpage 

\clearpage
\newpage 

\addcontentsline{toc}{section}{Supplementary Figures}

\section*{Supplementary Figures}
\label{sec:figs_SI}

\begin{figure*}[h]
\centering
\includegraphics[width=0.8\linewidth]{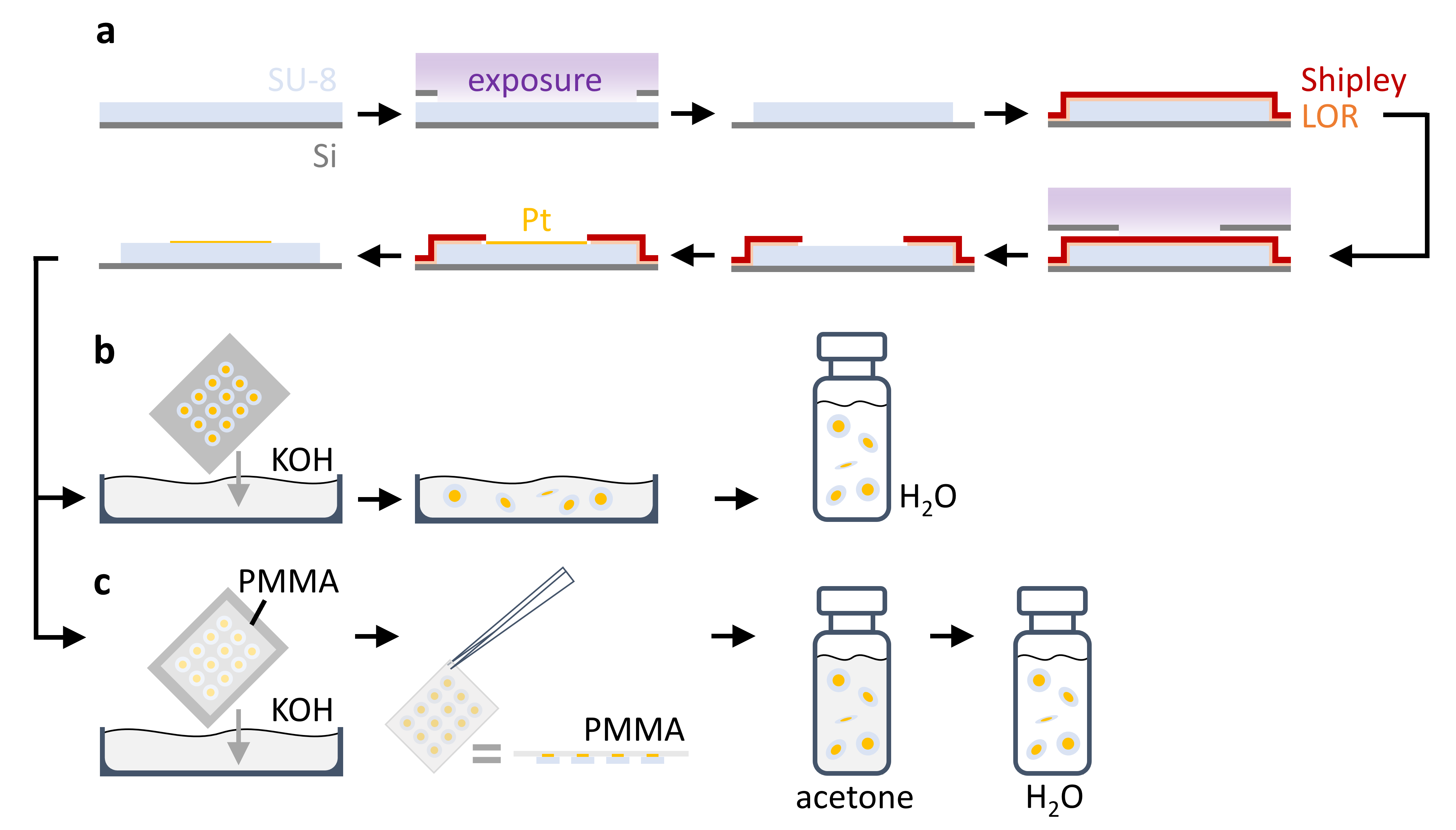}
\caption{\textbf{Beating particle fabrication steps.}
\textbf{a}, An array of SU-8 polymeric microdiscs were defined and patterned on a Si wafer with standard photolithography, followed by electron-beam physical vapor deposition of Pt on top. 
\textbf{b}, The particles were subsequently lifted off in heated KOH solution which etched into the Si substrate. The KOH was displaced by water in which the lifted off microparticles were stored.
\textbf{c}, Alternatively, a film of PMMA polymer was spun over the microparticle array. Together they would delaminate from the substrate in heated KOH solution. The PMMA was then removed with an acetone rinse. The lifted off particles were transferred to water for storage.}
\label{fig:fabrication}
\end{figure*}

\begin{figure*}[h]
    \centering
    \includegraphics[width=0.8\linewidth]{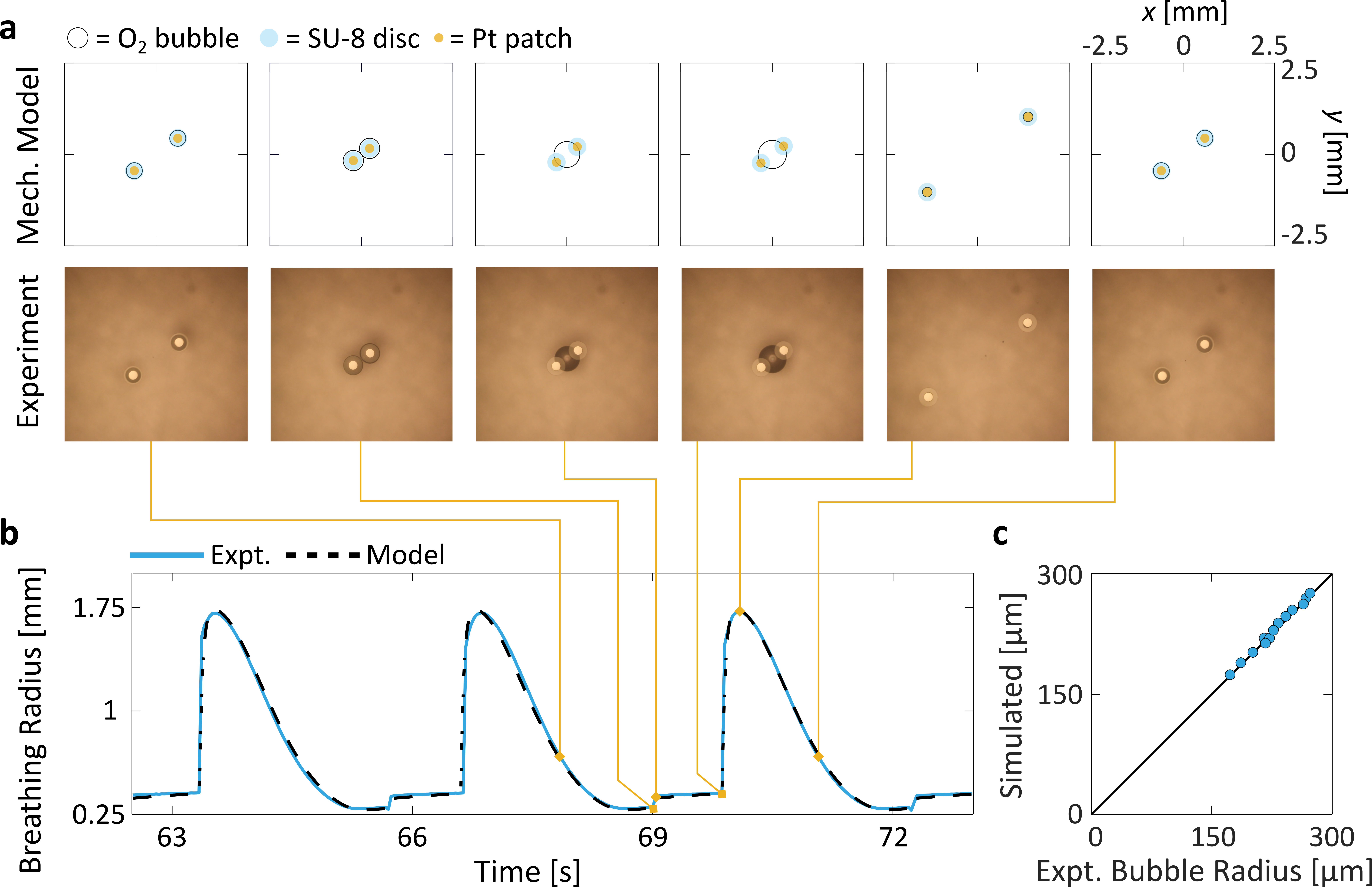}
    \caption{\textbf{Detailed comparison between experimental and simulated beating behaviours of two particles.} \textbf{a,} Mechanistic model simulations and experimental snapshots taken at representative stages of a beating cycle. \textbf{b,} The simulation and experiments are in excellent agreement, evident from the matching curves of the breathing radius, previously also shown in Fig. 1g. We note that the mechanistic model captures fine details of the self-oscillation, such as the subtle step change in (\textbf{b}) at approximately 69s. The step increase was a result of the merged bubble pushing the particles outwards slightly, reflected by both the experiment and simulation in (\textbf{a}). \textbf{c,} This panel shows the excellent agreement between the experimental bubble radii and those predicted by the mechanistic model. The former were measured manually from the raw video data.} 
    \label{fig:2pdetails_SI}
\end{figure*}

\begin{figure*}[h]
    \centering
    \includegraphics[width=0.6\linewidth]{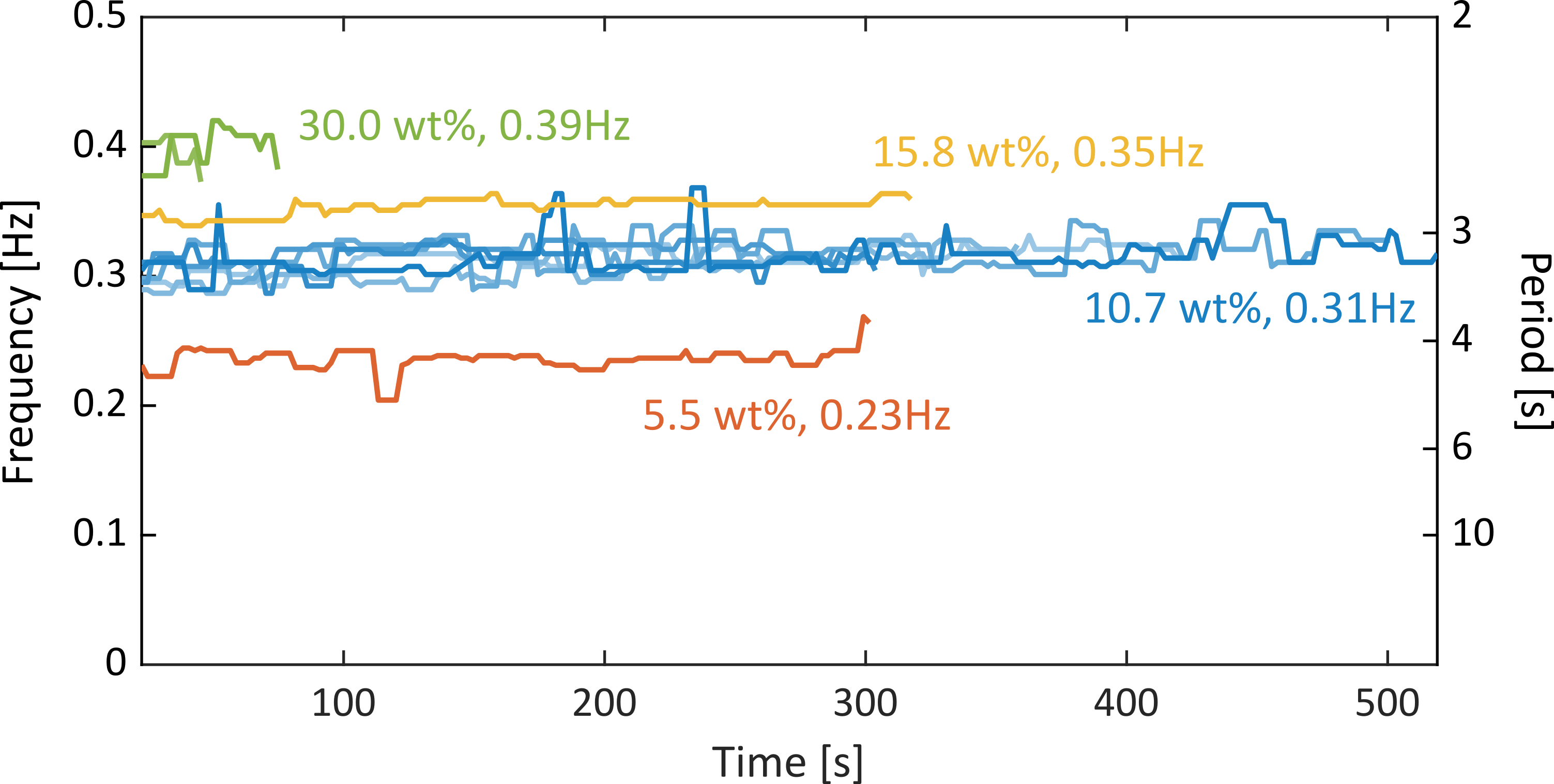}
    \caption{\textbf{Beating frequencies over time from moving window recurrence analyses.} The same histograms as in Fig. 1i of the main text were generated, but here only for breathing radius data within a moving window of 150 frames (5s). Frequencies calculated from the most probable recurrence time of each window were plotted as a function of time. The beating frequencies in all experiments are constant throughout, demonstrating robust periodicity. Furthermore, curves from experimental replicates overlap. The frequencies from moving window analyses agree with those shown in Fig. 1j for all the H$_2$O$_2$ concentrations. These concentrations respectively correspond to 6-, 3-, 2-, and 1-fold volumetric dilution of a 30wt\% H$_2$O$_2$ solution.} 
    \label{fig:windowed_recurrence}
\end{figure*}

\begin{figure*}[h]
    \centering
    \includegraphics[width=0.85\linewidth]{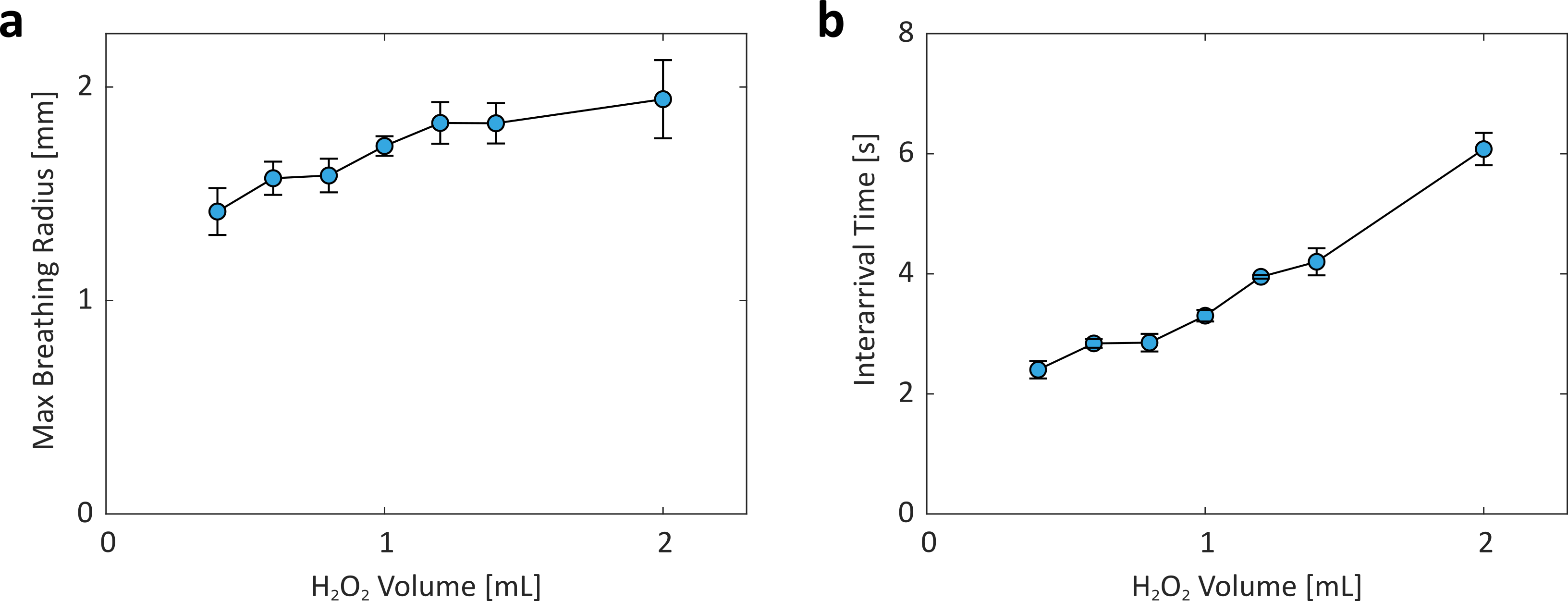}
    \caption{\textbf{Maximum breathing radius and interarrival time of two identical particles as a function of the H$_2$O$_2$ volume.}  A larger volume of H$_2$O$_2$ solution corresponds to a reduced curvature of the liquid-air interface the particles reside in, which in turn weakens the global restorational force that resists parting of the particles. The breathing radius (\textbf{a}) therefore increases with the H$_2$O$_2$ volume, which consequently lengthens the intervals between consecutive bubble collapses (\textbf{b}). Due to the periodicity of all these 2-particle systems, the respective interarrival times are equivalent to the periods of oscillation. Each error bar denotes a standard deviation among the oscillation cycles within an experiment.} 
    \label{fig:h2o2_volumes}
\end{figure*}

\begin{figure*}[h]
    \centering
    \includegraphics[width=0.85\linewidth]{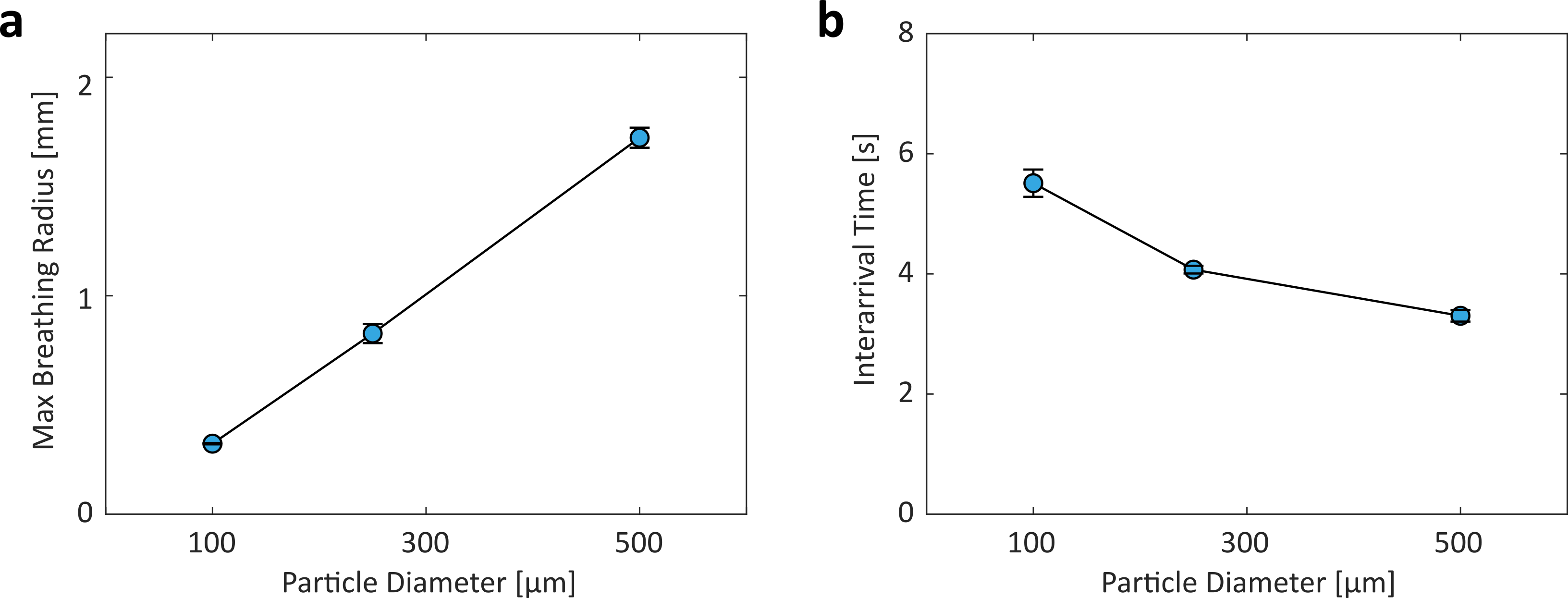}
    \caption{\textbf{Maximum breathing radius and interarrival time of two identical particles as a function of the particle size.} All particles were fabricated by depositing 5nm Cr and 50nm Pt onto 10$\mu$m-thick SU-8 polymer. The 500$\mu$m, 250$\mu$m, and 100$\mu$m-diameter particles were designed to have Pt patches 250$\mu$m, 125$\mu$m, and 100$\mu$m in diameter, respectively. As with Supplementary Figure \ref{fig:h2o2_volumes}, the oscillation amplitude exhibits an increasing trend with respect to the particle diameters (\textbf{a}). The interarrival times (\textbf{b}), however, decrease with the particle size. Due to the periodicity of all these 2-particle systems, the respective interarrival times are equivalent to the periods of oscillation. Each error bar denotes a standard deviation among the oscillation cycles within an experiment.} 
    \label{fig:particle_sizes}
\end{figure*}

\begin{figure*}[h]
    \centering
    \includegraphics[width=0.85\linewidth]{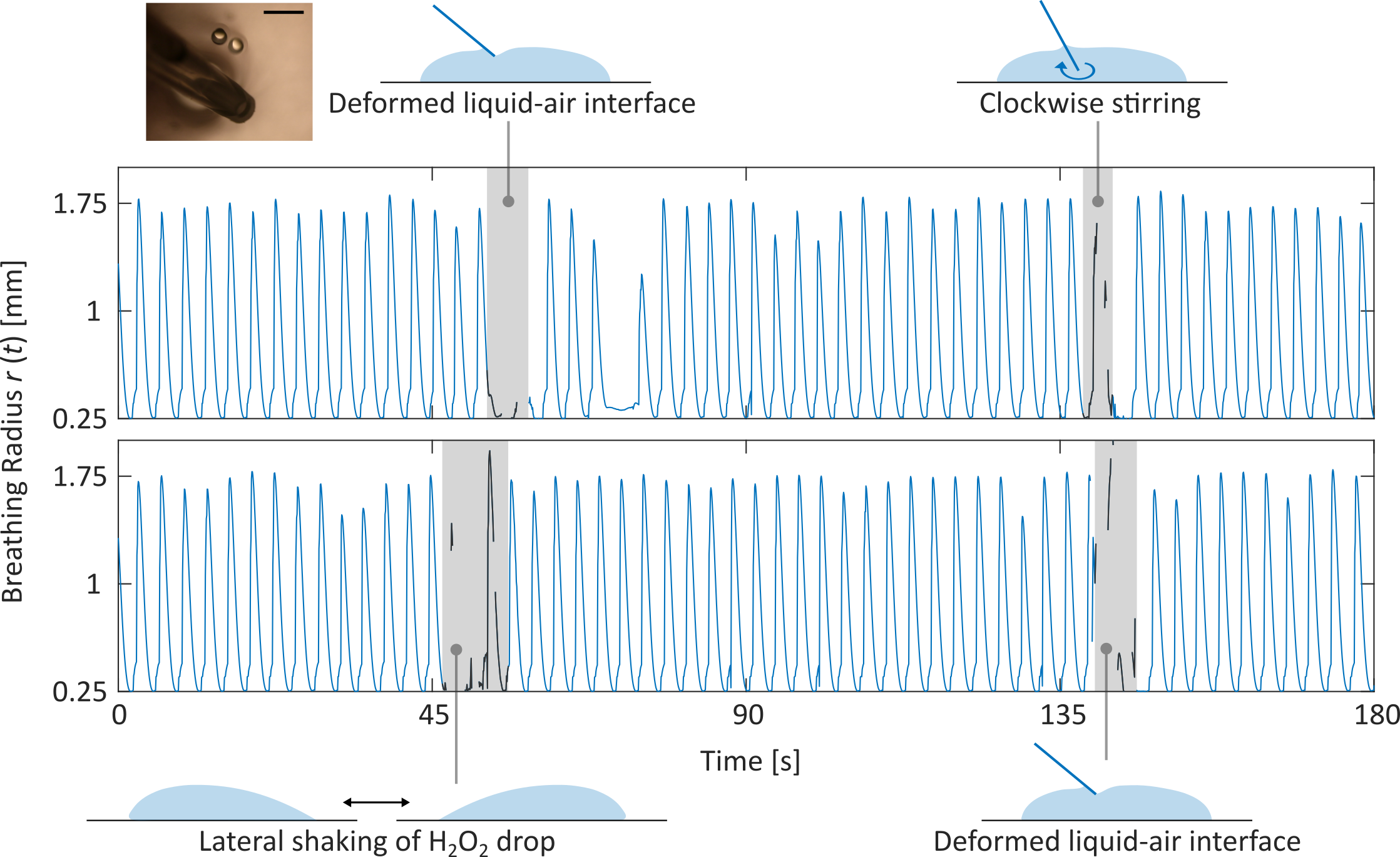}
    \caption{\textbf{Robustness of the emergent oscillation to perturbations.} In these two experiments, we intentionally disturbed a system of two identical particles by (i) deforming the liquid-air interface with a pipette~\cite{Solovev2010_microstrider}, (ii) stirring the H$_2$O$_2$ drop, and (iii) shaking the drop back and forth. It is evident in the breathing radius trajectories that the collective oscillation resumes promptly following the perturbations (shaded region) with its amplitude and periodicity unchanged, thus demonstrating robustness. Data discontinuities during the perturbations are a result of blurry frames or particles temporarily exiting the camera field-of-view. The inset micrograph shows the particles approaching the pipette due to the deformed interface. Scale bar, 1mm.} 
    \label{fig:robustness}
\end{figure*}

\begin{figure*}[h]
    \centering
    \includegraphics[width=0.65\linewidth]{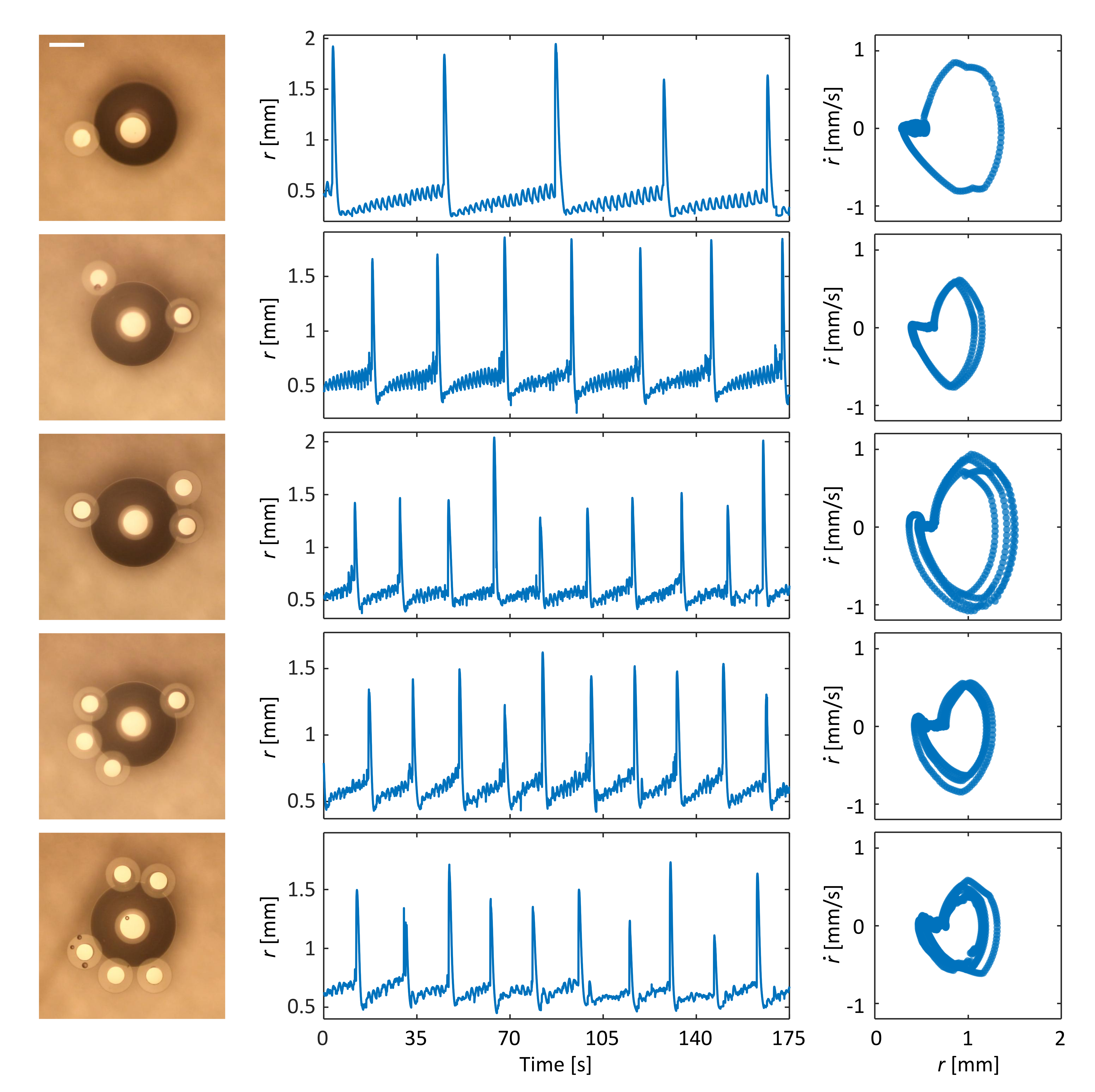}
    \caption{\textbf{Compiled snapshots, breathing radius trajectories, and phase portraits for heterogeneous/DL systems of \textit{N} = 2 to 6.} Systems of all sizes exhibited clear periodicity in their beating behaviours with stable limit cycles. Scale bar, 500$\mu$m. All experiments were performed in 1mL of 10.7wt\% H$_2$O$_2$.} 
    \label{fig:master_l}
\end{figure*}

\begin{figure*}[h]
    \centering
    \includegraphics[width=0.65\linewidth]{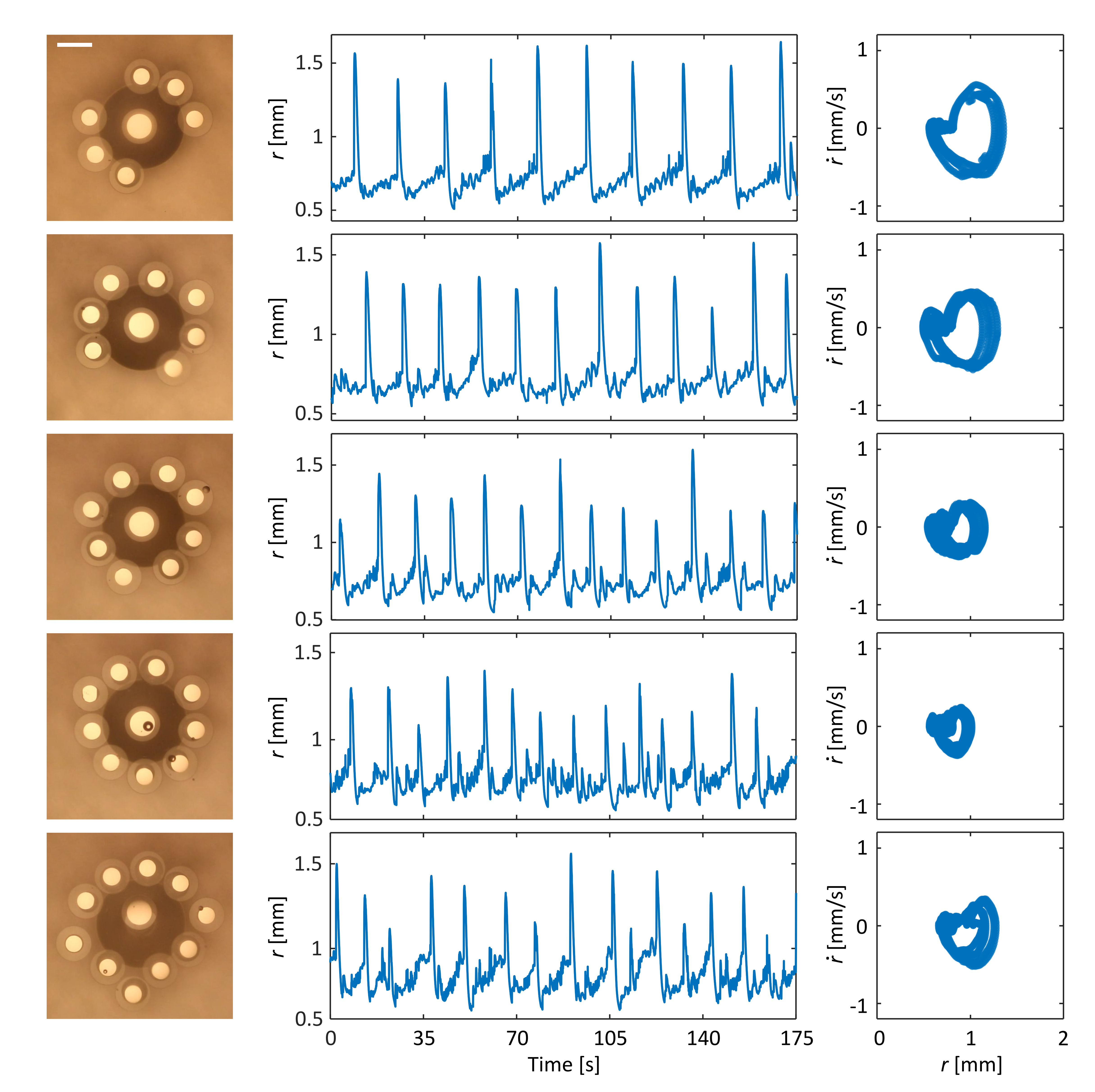}
    \caption{\textbf{Compiled snapshots, breathing radius trajectories, and phase portraits for heterogeneous/DL systems of \textit{N} = 7 to 11.} Systems of all sizes exhibited clear periodicity in their beating behaviours with stable limit cycles. Scale bar, 500$\mu$m. All experiments were performed in 1mL of 10.7wt\% H$_2$O$_2$.} 
    \label{fig:master_r}
\end{figure*}

\begin{figure*}[h]
    \centering
    \includegraphics[width=0.65\linewidth]{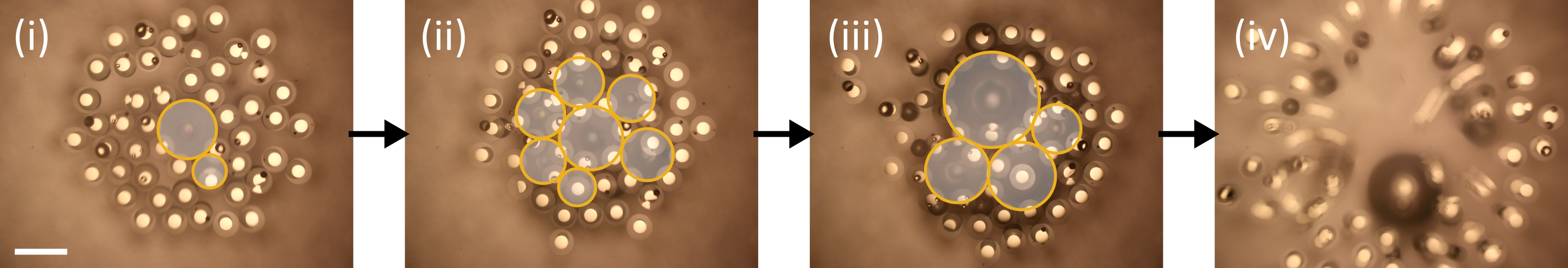}
    \caption{\textbf{Progression of a large-scale homogeneous collection.} While we have established in Fig. 2 of the main text that the periodicity of homogeneous systems breaks down easily as \textit{N} increases, we observe intriguing hierarchical organization of the bubbles in this 50-particle collective over a span of 8s. Bubbles from individual particles merge and grow (i), resulting in the intermediate situation in (ii) where a large bubble situated at the H$_2$O$_2$ drop's apex is packed around by smaller ones. Following further merger and growth (iii), the system eventually collapses (iv). Highlighted in yellow circles are bubbles larger than 350$\mu$m in radius. With a number of particles distributed along the perimeter, a bubble is observed to grow far beyond the typical threshold size in few-particle homogeneous systems (\textit{cf.} bubble sizes in DL systems in Supplementary Figs.~\ref{fig:master_l} and~\ref{fig:master_r}). Scale bar, 1mm. All experiments were performed in 1mL of 10.7wt\% H$_2$O$_2$.} 
    \label{fig:many_particles}
\end{figure*}

\begin{figure*}[h]
    \centering
    \includegraphics[width=0.4\linewidth]{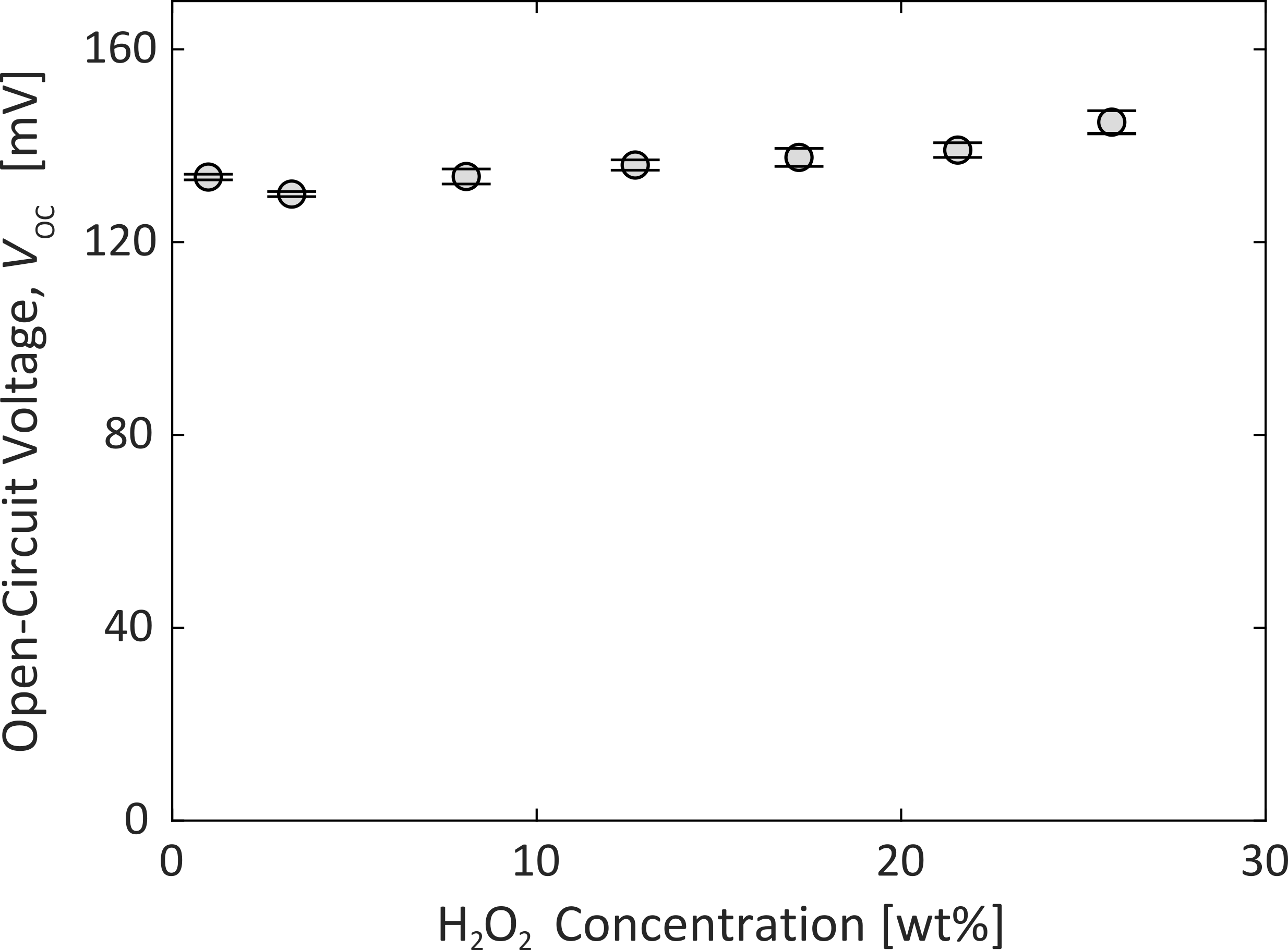}
    \caption{\textbf{Open-circuit voltage of a Pt-Ru device as a weak function of H$_2$O$_2$ concentration.} The observation is explained by the auto-redox nature of the H$_2$O$_2$ decomposition reaction (Supplementary Section 4). Error bar, standard deviation.} 
    \label{fig:ocv_SI}
\end{figure*}

\begin{figure*}[h]
    \centering
    \includegraphics[width=0.4\linewidth]{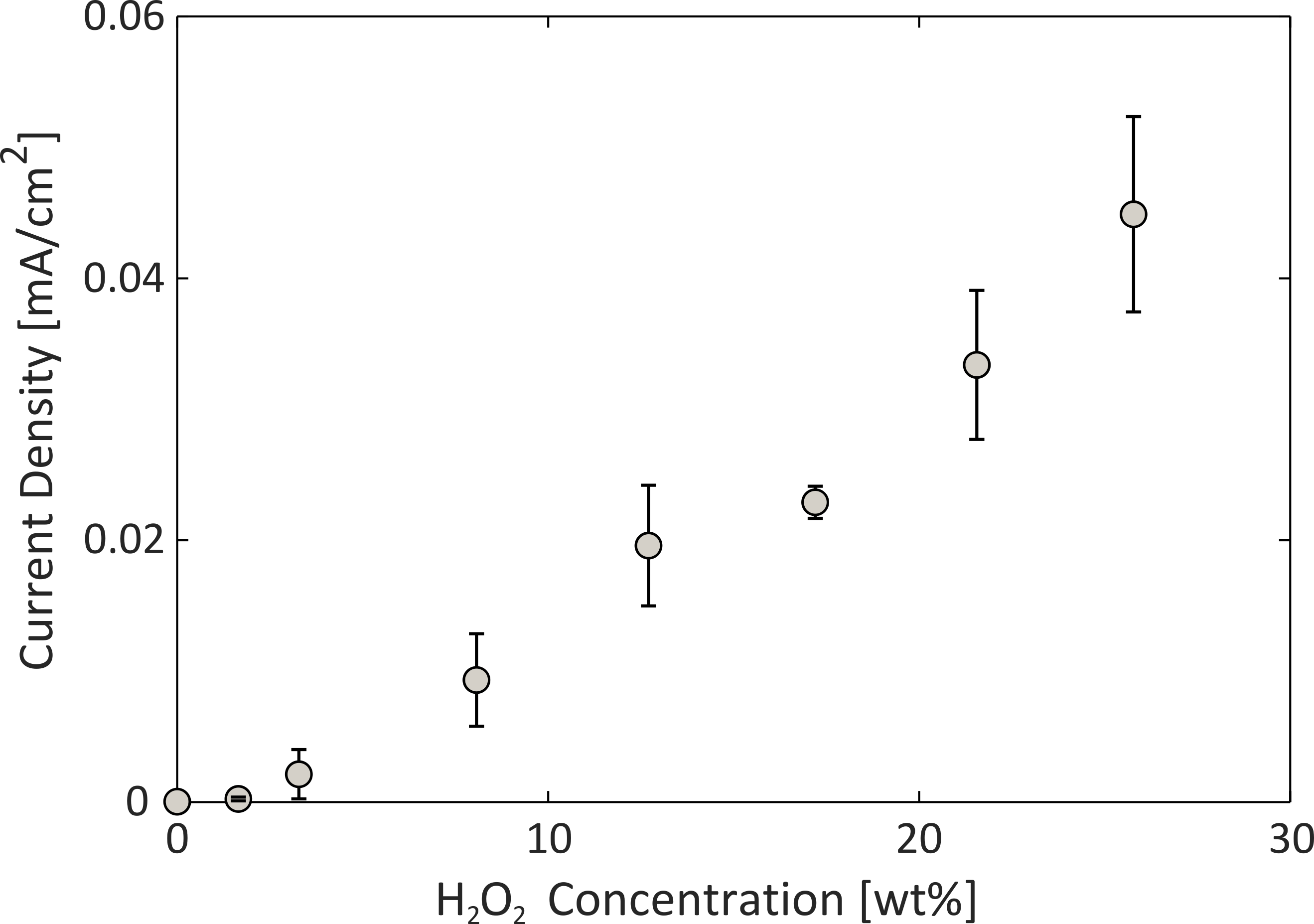}
    \caption{\textbf{Short-circuit current density as a function of H$_2$O$_2$ concentration for a Pt-Au device.} (\textit{cf}. Fig. 4c of the main text). Error bar, standard deviation.} 
    \label{fig:currentPtAu_SI}
\end{figure*}

\begin{figure*}[h]
\centering
\includegraphics[width=0.65\linewidth]{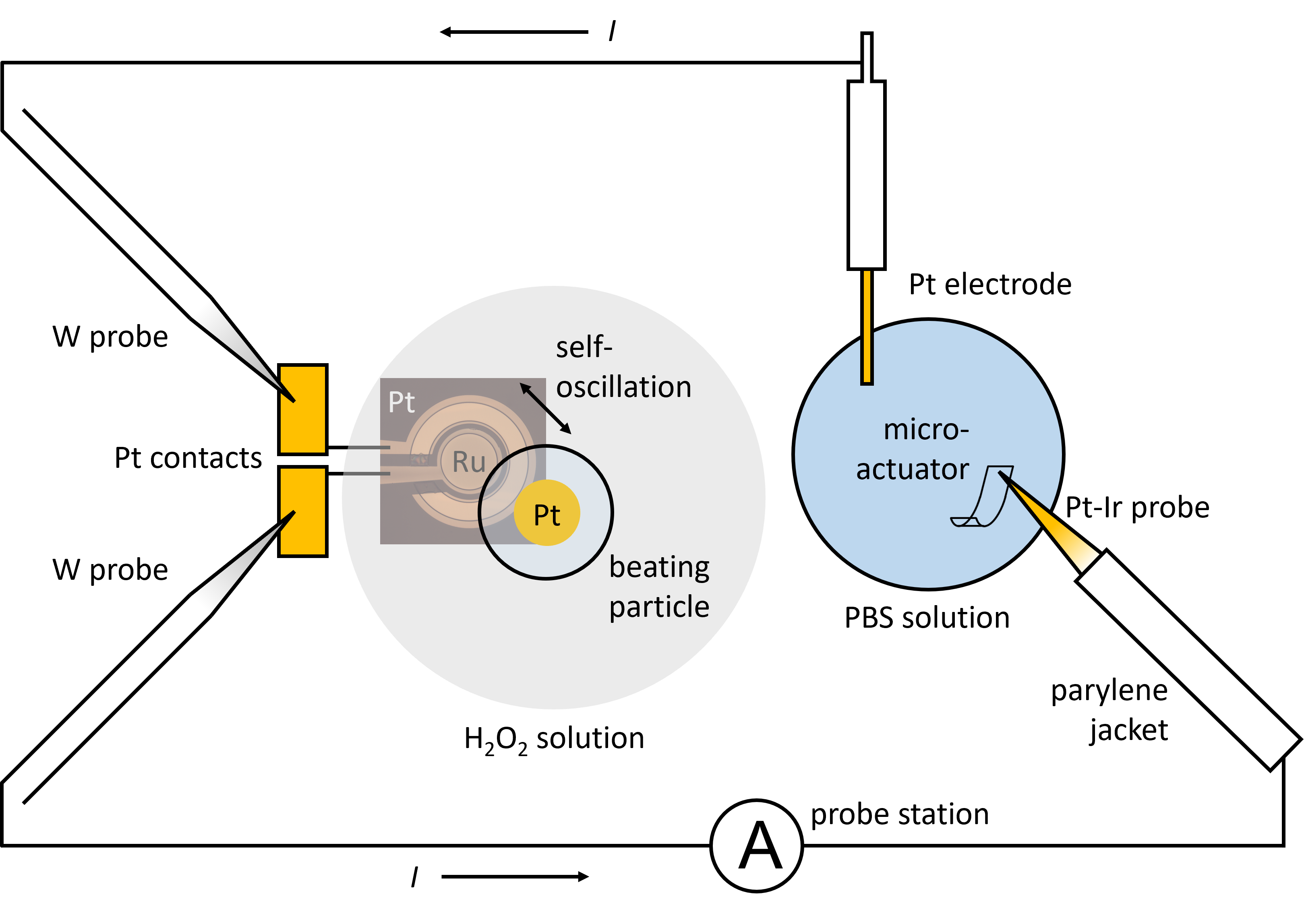}
\caption{\textbf{Bimorph actuator experimental setup.} The microactuator in PBS solution is connected via external wiring to the beating system in an H$_2$O$_2$ drop. The mechanical self-oscillation is translated to an oscillatory electrical current as illustrated in Fig.~4a, which powers cyclic motion of the actuator (Fig.~4e).}
\label{fig:actuator_setup}
\end{figure*}

\begin{figure*}[h]
    \centering
    \includegraphics[width=0.5\linewidth]{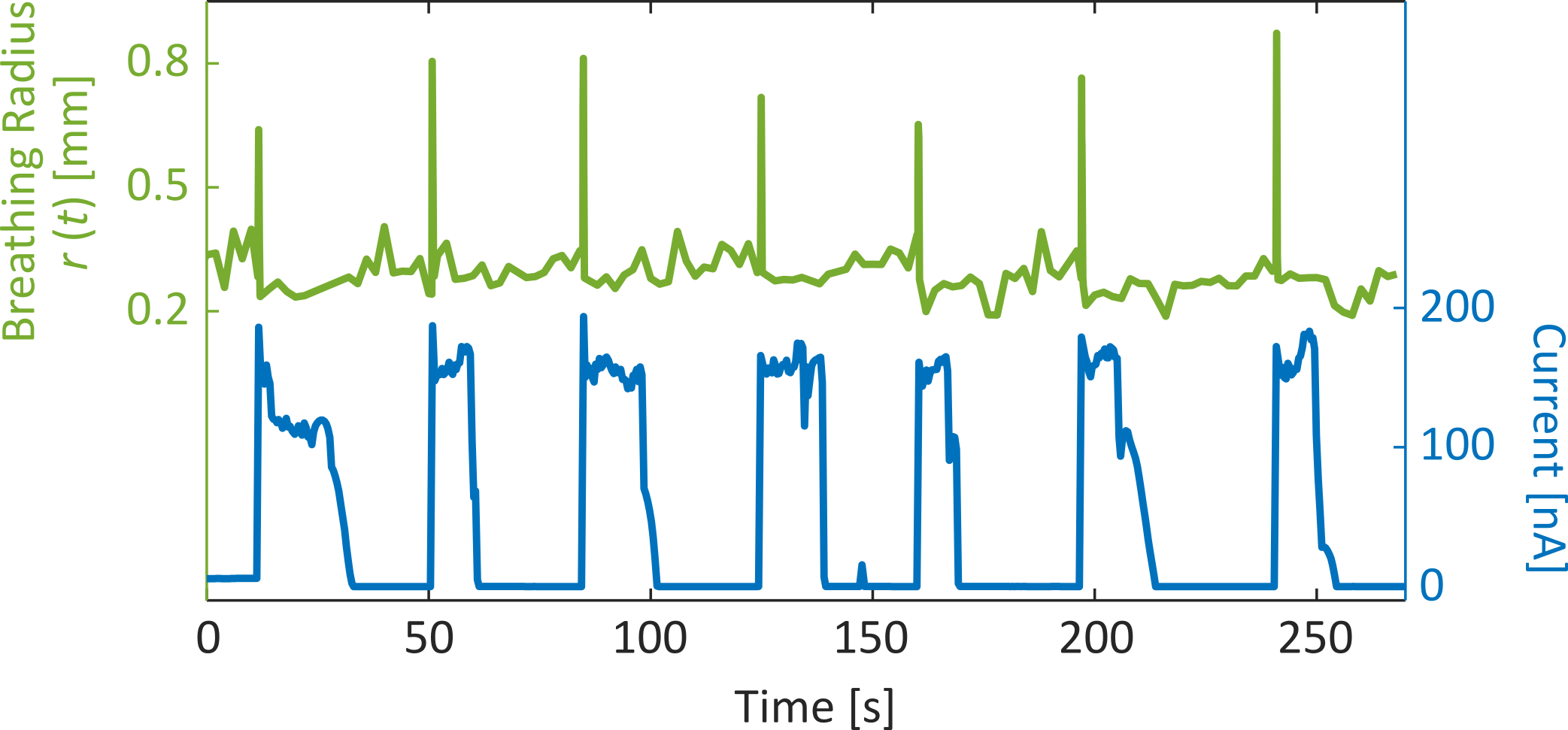}
    \caption{\textbf{Oscillatory mechanical beating drives on-board oscillatory current.} (See also Fig. 4e of the main text). As a standard 500-$\mu$m particle beats with a Pt-Ru fuel cell device (Fig. 4b, also Methods), the bubbles collapse at regular intervals as indicated by the spikes in the breathing radius trajectory ($r(t)$, top). Removal of the bubbles restores the electrochemical reactivity of the fuel cell electrodes, and therefore the current (bottom) peaks precisely as $r(t)$ does. The current measured in this experiment is an order of magnitude higher than that in Fig. 4e since the system characterized here was not connected to an actuator.} 
    \label{fig:currentPtAu_SI}
\end{figure*}

\clearpage
\newpage 

\addcontentsline{toc}{section}{Supplementary References}

\printbibliography[title=Supplementary References]